\documentclass[final]{IEEEtran} 
\usepackage{fancyhdr}
\usepackage{epsfig}
\usepackage{threeparttable}
\usepackage{epsf,epsfig}
\usepackage{amsthm}
\usepackage{amsmath}
\usepackage{amssymb}
\usepackage{amsfonts}
\usepackage[noadjust]{cite}
\usepackage{dsfont}
\usepackage{subfigure}
\usepackage{color}
\usepackage{url}
\usepackage{mathrsfs}

\newlength{\eqboxstorage} \newcommand{\eqbox}[1]{
\setlength{\eqboxstorage}{\fboxsep} \setlength{\fboxsep}{6pt} \boxed{#1} \setlength{\fboxsep}{\eqboxstorage}}

\newtheorem{theorem}{Theorem}

\newtheorem{definition}{Definition}

\newtheorem{lemma}{Lemma}

\newtheorem{assumption}{Assumption}

\def\E{\mathsf{E}}

\def\phi{\varphi}

\def\l{\left}
\def\r{\right}
\def\({\left(}
\def\){\right)}

\def\bff{{\mathbf{f}}}

\def\bh{{\mathbf{h}}}

\def\bq{{\mathbf{q}}}

\def\b0{{\mathbf{0}}}

\newcommand{\Pout}{P_{\mathsf{out}}}

\newcommand{\nn}{\nonumber}

\begin{document}
\title{ An Analytical Framework for Multi-Cell Cooperation via Stochastic Geometry and Large Deviations}
\author{Kaibin Huang and Jeffrey G. Andrews \thanks{K. Huang is with the Hong Kong Polytechnic University, Hong Kong  and J. G. Andrews is with The University of Texas at Austin, TX.  Email: huangkb@ieee.org, jandrews@ece.utexas.edu. This paper has been presented in part at IEEE Globecom 2011 and IEEE Intl. Conf. on Communications 2012. Updated on \today.  }}

\maketitle
\begin{abstract}
Multi-cell cooperation (MCC) is  an approach for mitigating inter-cell interference in dense cellular networks. Existing studies on MCC performance typically rely on either over-simplified Wyner-type models or complex system-level simulations. The promising theoretical results (typically using Wyner models) seem to materialize neither in  complex simulations nor  in practice.  To more accurately investigate the theoretical performance of MCC, this  paper models an entire plane of interfering cells as a Poisson random tessellation. The base stations (BSs) are then clustered using a regular lattice, whereby BSs in the same cluster mitigate mutual interference by beamforming with perfect channel state information. Techniques from stochastic geometry and large-deviation theory  are applied to analyze the outage probability as a function of the mobile locations, scattering environment and the average number of cooperating BSs per cluster, $\ell$.  For mobiles near the centers of BS clusters,  it is shown that outage probability  diminishes  as $O(e^{- \ell^{\nu_1}})$ with $0\leq \nu_1\leq 1$ if scattering is sparse,  and as $O(\ell^{-\nu_2})$ with $\nu_2$ proportional to the signal diversity order if scattering is rich.  For randomly located mobiles, regardless of scattering, outage probability is shown to scale as $O(\ell^{-\nu_3})$ with $0\leq \nu_3 \leq 0.5$.   These results confirm analytically that cluster-edge mobiles are the bottleneck for network coverage and provide a plausible analytic framework for more realistic analysis of other multi-cell techniques.

\end{abstract}

\section{Introduction}
Inter-cell interference limits the performance of cellular downlink networks but can be  suppressed by multi-cell cooperation (MCC). The existing high-speed backhaul links allow base stations (BSs) to  exchange data and channel state information (CSI). { Thereby,  cells can be  grouped into finite clusters  and BSs in a same cluster cooperate to decouple the assigned  mobiles  \cite{Lozano:FundamentalLimitCooperation:2012, KarakFoschini:NeworkCoordinationCellular:2006, DahYu:CoordBeamformMulticell:2010, Simeone:LocalBSCooperationFinite-CapLink:2009}.}  Despite extensive research  conducted on MCC,  the fundamental limits of cellular networks with MCC remain largely unknown due to the lack of an accurate and yet tractable  network model. This paper 
addresses this issue by proposing   a  novel model  constructed using  a Poisson point process (PPP) for BSs and a hexagonal lattice for clustering said BSs. Based on  this model, techniques from stochastic geometry and large-deviation theory   are applied to quantify the relation between network coverage and the average number of cooperating BSs. 

\subsection{Modeling Multi-Cell Cooperation}

Quantifying the performance gain by  MCC requires accurately modeling 
the cellular-network architecture and { accounting  for the relative locations of  BSs and mobiles.} These factors  are barely modeled   in    Wyner-type  models where base stations are arranged in a line or circle, interference exists only between neighboring cells and path loss is  represented by a fixed scaling factor \cite{Shamai:InfoTheoSymmetricCellularChannel:2002}. Due to their tractability, Wyner-type models are commonly used in  information-theoretic studies of MCC \cite{Shamai:CellularDLCoProcessing:2001, SimeoneShamai:DownlinkMulticellLimitedBackhaul:2009, Simeone:LocalBSCooperationFinite-CapLink:2009}, but fail  to account for mobiles' random locations \cite{XuAndrews:AccuracyWynerModel} and finite BS clusters in practice  due to a constraint on the cooperation overhead \cite{Lozano:FundamentalLimitCooperation:2012, PapadGesbert:DynamicClusteringMulticellCoop:2008, Zhang:NetworkedMIMOClusterLinearPrecod:2009}. The traditional hexagonal-grid model provides a better approximation of a practical cellular network, however,  at the cost of tractability \cite{Foschini:CoordinatingMultiAntennaCellularNetworks:2006}. An alternative modeling approach  is to model  BSs using a PPP and construct cells as a  random spatial tessellation \cite{Andrews:TractableApproachCoverageCellular:2010}. The random  model  captures cell irregularity, is about as accurate as the hexagonal-grid model, and allows  analysis  using stochastic geometry { \cite{HaenggiAndrews:StochasticGeometryRandomGraphWirelessNetworks, Win:MathTheoryNetInterference:2009}}. 

Building on \cite{Andrews:TractableApproachCoverageCellular:2010} which assumes  single-cell transmission,  in this paper BSs are modeled as a homogeneous PPP that partitions the horizontal plane into Voronoi cells.  Mobiles in each cell are randomly located and time share  the corresponding BS. BSs are then  clustered  using a larger hexagonal lattice \footnote{{ The hexagonal lattice is chosen arbitrarily for exposition. It is  straightforward to extend the current analysis to BS clustering using other  types of regular lattice or 
random spatial   tessellations by modifying the definitions of the variables  $\rho, \tilde{\rho}$ and $D$ (defined in the sequel) based on the cell geometry. 
} } to cooperate { by \emph{interference coordination}  where BSs in the same cluster mitigate interference to each others' mobiles by zero-forcing beamforming that also  achieves   transmit-diversity  gain \cite{Gesbert:MultiCellMIMOCooperativeNetworks:2010}}.  Furthermore, to cope with fading, channel inversion is applied such that received signal power is fixed. { This scheme is considered for simplifying analysis and can be implemented in practice by combining a transmit-diversity technique and automatic gain control widely used in code-division-multiple-access systems. It is worth mentioning that channel inversion is found in this research to reduce outage probability compared with fixed-power transmission.  }
Outage probability specifies  the fraction of mobiles outside network coverage for a target signal-to-interference ratio (SIR), assuming an interference limited network. This is the case of interest for MCC and of operational relevance for cellular networks.
 Let the average  number of BSs in a cluster be denoted as $\ell$, called the \emph{expected BS-cluster size}. 
This paper focuses on quantifying the asymptotic rate at which outage probability diminishes as $\ell$ increases. 

{This and any other clustering methods with finite cluster sizes and only intra-cluster cooperation have the drawback  of cluster-edge mobiles exposed to strong inter-cluster interference as quantified in the subsequent analysis. Intuitively, a better approach is to allow overlapping BS clusters for protecting cluster-edge mobiles. BS cooperation based on this approach can be implemented efficiently using belief propagation and message passing \cite{NgHanly:DownlinkCoopBS:2008, RanganMadan:BeliefPropagationInterferenceCoord:2012, Sohn:BeliefPropDistributedBeamCoop:2011} but will eventually involve all BSs in the network and cause potential issues including overwhelming backhaul overhead, excessive delay and network instability. For these reasons, BS clusters  in practice  are usually disjoint \cite{Irmer:CoMPConceptsPerformFieldTrial:2011}. This investigation suggests a much simpler approach for suppressing inter-cluster interference for  cluster-edge mobiles by combining the current method of BS clustering with \emph{fractional frequency reuse}  \cite{Boudreau:InterfCoordCancel4G:2009} along cluster edges as discussed in the sequel.}

There exists a rich literature on analyzing outage probability for wireless networks with Poisson distributed transmitters { \cite{PintoWin:CommPoisson:2010, GantiHaenggi:OutageClusteredMANET:2009, WeberAndrews:TransCapWlssAdHocNetwkOutage:2005, ChanHanly:CalOutageCDMANetwkSpatialPoission:2001}}. Given that outage probability has no closed-form expressions \cite{Baccelli:AlohaProtocolMultihopMANET:2006,Lowen:PowerLawShotNoise:1990}, a common analytical approach is to derive bounds on outage probability using  probabilistic inequalities \cite{WeberAndrews:TransCapAdHocNetwkDistSch:2006}, which are sufficiently simple and tight for evaluating network performance given specific   transmission techniques e.g.,  bandwidth partitioning \cite{JindalAndrews:BandwidthPartitioning:2007} and multi-antenna techniques { \cite{VazeHeath:TransCapacityMultipleAntennaAdHocNetwork, LouieMacKay:SpatialMultiplexDiversityAdHocNetwork}}. The accuracy of these outage-probability bounds requires the presence  of strong interferers for mobiles.   Similar bounds for cellular networks with MCC can be loose since  interference is suppressed using MCC. Therefore,  this work deploys an alternative approach where large-deviation theory \cite{DemboBoo:LargeDeviation} is applied to quantify  the exponential decay  of outage probability as $\ell\rightarrow\infty$. A similar approach was applied in \cite{Ganesh:LargeDeviationInterferenceWirelessNet} to analyze    the  tail probability of interference  in a wireless ad hoc network.

\subsection{Summary of Contributions and Organization}
To apply techniques from large-deviation theory, a new performance metric called the  \emph{outage-probability exponent} (OPE) is defined as follows. 
Since  the network is interference limited and hence noise is negligible,   the outage probability for an arbitrary mobile, denoted as $\Pout$, is given as 
\begin{equation}\label{Eq:Pout:Def}
\Pout = \Pr\l(\frac{\omega}{I} < \theta \r)
\end{equation}
where $\omega$ and $I$ represent the fixed received  signal power and random interference power, respectively, and $\theta > 0$ is the outage threshold. Then the OPE is defined as 
\begin{align}
\phi(\ell) &= -\log \Pout\label{Eq:OPE:Def:a}\\
& = -\log \Pr\l(I >  \frac{\omega}{\theta} \r)   \label{Eq:OPE:Def}
\end{align}
where $\Pout$ and $I$ are functions of $\ell$ with $\ell$  omitted for ease of notation.   It follows that deriving the scaling of $\phi(\ell)$ as $\ell\rightarrow\infty$ yields the exponential decay rate of $\Pout$. Using large-deviation theory, simple OPE scalings are  derived for different network configurations based on  the  rates at which the tail probabilities of random network parameters diminish as $\ell\rightarrow\infty$. 

With interference being suppressed by increasing $\ell$, the network will eventually operate in the noise limited regime,  for which the outage-probability for a typical   mobile  is either zero or one depending on if the received signal-to-noise ratio  $\omega/\sigma^2$ is below or above $\theta$.  The value of $\omega$ depends on the average transmission power of BSs and channel distribution [see \eqref{Eq:PwrCtrl:RxPwr} in the sequel]. Therefore, the OPE becomes  irrelevant for the case of a noise-limited network with channel inversion.

\begin{table*}[t]
\caption{Summary of notation}
\begin{center}
\begin{tabular}{|c|l|}
\hline
{\bf Symbol}  &   {\bf Meaning}\tabularnewline 
\hline
$\phi$, $\phi^{cc}$ &OPE for a (typical, cluster-center) mobile  \\
\hline
$I$, $I^{cc}$  & Received interference power for a (typical, cluster-center) mobile \\
\hline
$\ell$  & Expected BS-cluster size\\
\hline
$M$ & Number of BSs in a typical cluster\\
\hline
$\Phi$, $\lambda$ & PPP of BSs, density of $\Phi$\\
\hline
$\Omega$ & Hexagonal lattice for clustering BSs\\
\hline
$Y^*, T^*, U^*$ & Typical BS, BS-cluster center and mobile\\
\hline
$\mathcal{U}^*$ &  Cluster of mobiles served by the typical BS cluster\\
 \hline
$\mathcal{C}(T, r)$ & Hexagon centered at $T\in\mathds{R}^2$ and having the distance  $r$ from $T$ to the boundary\\
\hline
$\rho$, $\tilde{\rho}$  & Distance from the center of a cluster region to an (edge, vertex)\\ 
\hline
$u(Y)$ &Mobile served by BS $Y$\\
\hline
$L_Y$ &Distance from BS $Y$ to the affiliated mobile\\
\hline
$P_Y$ &Transmission power for  BS $Y$\\
\hline
$\bff_Y$ & Beamformer used at   BS $Y$\\
\hline
$\bh_{UY}$ & Vector channel from  BS $Y$ to  mobile $U$\\
\hline
$\alpha$ & Path-loss exponent\\
\hline
$\theta$ & Outage threshold \\
\hline
$\omega$ & Fixed received signal power at a mobile \\
\hline 
$N$, $\nu$ & Signal diversity order for a typical mobile, the minimum value of $N$ \\
\hline
$D$ & Distance from a typical mobile to the boundary of the corresponding cluster \\
\hline
\end{tabular}
\end{center}
\label{Table}
\end{table*}

The main contributions of this paper are summarized as follows. 
\begin{enumerate}

\item  Consider a mobile located at the center of an arbitrary BS cluster, called a \emph{cluster-center} mobile, and  sparse scattering where beams have bounded amplitudes.  Given MCC, the OPE for a cluster-center mobile, denoted as $\phi^{cc}$,  is shown to scale \footnote{Two functions $f(z)$ and $g(z)$ are asymptotically equivalent if $\frac{f(z)}{g(z)}\rightarrow 1$ as $z\rightarrow \infty$, denoted as $f(z) \sim g(z)$; the cases of $\lim_{z\rightarrow\infty}\frac{f(z)}{g(z)}\geq 1$ and $\lim_{z\rightarrow\infty}\frac{f(z)}{g(z)}\leq 1$ are represented by $f(z) \succeq g(z)$ and $f(z) \preceq g(z)$, respectively.} as follows: 
\begin{enumerate}
\item for the path-loss exponent $\alpha > 4$, 
\begin{equation}
c_1 \ell \preceq\phi^{cc}(\ell) \preceq  \frac{4c_1}{3} \ell, \qquad \ell \rightarrow\infty; \nn
\end{equation}
\item for   $2 < \alpha \leq 4$,  
\begin{equation}
c_2 \ell^{\frac{\alpha}{4}}  \preceq \phi^{cc}(\ell) \preceq \frac{4c_1}{3} \ell, \qquad \ell \rightarrow\infty \nn
\end{equation}
where $c_1$ and $c_2$ are constants. 
\end{enumerate}
This result shows  that outage probability diminishes \emph{exponentially} as $\ell\rightarrow\infty$ for a high level of spatial separation  ($\alpha > 4$) or at least \emph{sub-exponentially} if the level is  moderate-to-low ($2 < \alpha \leq 4$).

\item  Consider a mobile with a randomly distributed location, called a \emph{typical mobile},   \footnote{A \emph{typical} point of a random point process is chosen from the process by uniform sampling such that all points are selected with equal probability.} and also MCC with sparse scattering. 
The scaling of the  corresponding  OPE is proved to be 
\begin{equation}\label{Intro:OPE:Typical}
\frac{1}{2}\l(1 - \frac{2}{\alpha}\r)\log \ell \preceq \phi(\ell)\preceq   \frac{1}{2}\log \ell, \qquad \ell \rightarrow \infty. 
\end{equation}
This result implies that outage probability decays as $\ell\rightarrow\infty$ following a \emph{power law} with an exponent smaller than $0.5$. This decay rate is much slower than the sub-exponential (up to exponential) rate  for a cluster-center mobile. The reason is that a typical mobile may lie near a cluster edge and consequently is exposed to strong inter-cluster interference. Comparing the outage-probability decay rates for cluster-center and typical mobiles suggests that cluster-edge  mobiles are the bottleneck of network coverage even with MCC and protecting them from inter-cluster interference (e.g., assigning   dedicated frequency channels) can significantly improve network coverage. 

\item Consider MCC with rich scattering modeled as Rayleigh fading. { Note that fading affects the interference distribution but not received signal power that is fixed given channel inversion.} 
The OPE for a cluster-center mobile is shown to satisfy 
\begin{equation}\nn
\l(\frac{1}{2} \alpha \nu -1\r)\log \ell\preceq \phi^{cc}(\ell) \preceq \frac{1}{2} \alpha \nu \log \ell, \qquad \ell\rightarrow\infty 
\end{equation}
where $\nu > 1$ is the minimum signal diversity order over different cells. It follows that outage probability decays as $\ell\rightarrow\infty$ following a \emph{power law} with an exponent approximately proportional to $\alpha$ and $\nu$. By comparing the outage-probability decay rates for sparse and rich scattering, it is found that { additional randomness in interference due to fading } degrades the reliability of communications near cluster centers significantly. 

\item Last, the OPE scaling for a typical mobile with sparse scattering from \eqref{Intro:OPE:Typical} is shown to also hold for a typical mobile with rich scattering. { The OPE scaling is  largely determined by the probability that the mobile lies near cluster boundaries and outside network coverage  due to  strong inter-cluster interference. As a result, the scaling is insensitive to if fading is present, which, however, impacts the OPE scaling for a cluster-center mobile. }

\end{enumerate}

The remainder  of the paper is organized as follows. The network model is  described  in Section~\ref{Section:Model}. The OPEs with sparse scattering and with rich scattering are  analyzed  in Section~\ref{Section:OPE:Sparse} and Section~\ref{Section:OPE:Rich}, respectively.  Simulation results are presented in Section~\ref{Section:Simulations} followed by concluding remarks  in Section~\ref{Section:Conclusion}. The appendix contains the proofs of lemmas.

\subsection{Notation}
The complement of a set $\mathcal{X}$ is represented by $\bar{\mathcal{X}}$. The operator $|X|$  on $X$ gives its cardinality if $X$ is a set or  the 
distance from $X$ to the origin if $X$ represents a point in the plane $\mathds{R}^2$.   The superscripts $T$ and $\dagger$ represent the matrix transpose and  Hermitian transpose operations, respectively.   

The families of distributions having 
\emph{regularly varying} and \emph{Weibull-like}  tails are represented respectively by $\mathsf{RV}(\tau)$ and $\mathsf{WE}(\tau)$ where $\tau > 0$ is the index, and defined  as follows. Define the distribution functions $\mathcal{F}$ and $\bar{\mathcal{F}}$ of a random variable (rv) $X$ as $\mathcal{F}(x) = \Pr(X \leq x)$ and $\bar{\mathcal{F}} = \Pr(X > x)$. 
 The rv  $X\in\mathsf{RV}(\tau) $ if  $\bar{\mathcal{F}}(x)  = x^{-\tau}\mathcal{P}(x)$ as $x \rightarrow\infty$ with $\mathcal{P}(x)$ being a \emph{slowly varying} function, namely $\lim_{x\rightarrow\infty}\frac{\mathcal{P}(t x)}{\mathcal{P}(x)}= 1$ for all $t > 0$ \cite{AsmussenBook:RuinProb}. If $X\in\mathsf{WE}(\tau)$,  $X$ has support $[0, \infty)$ and $\bar{F}(x)\sim cx^\eta e^{-\zeta x^\tau}$ form some $\beta \in \mathds{R}$, $\tau \in (0, 1)$, $\zeta > 0$ and  a constant $c > 0$ \cite{Mikosch:LargeDevHeavyTailedInsurance:1998}.

Other notation is summarized in Table~\ref{Table}.

\section{Network Model}\label{Section:Model}
\subsection{Network Architecture} 
\begin{figure*}[t]
\begin{center}
\includegraphics[width=15cm]{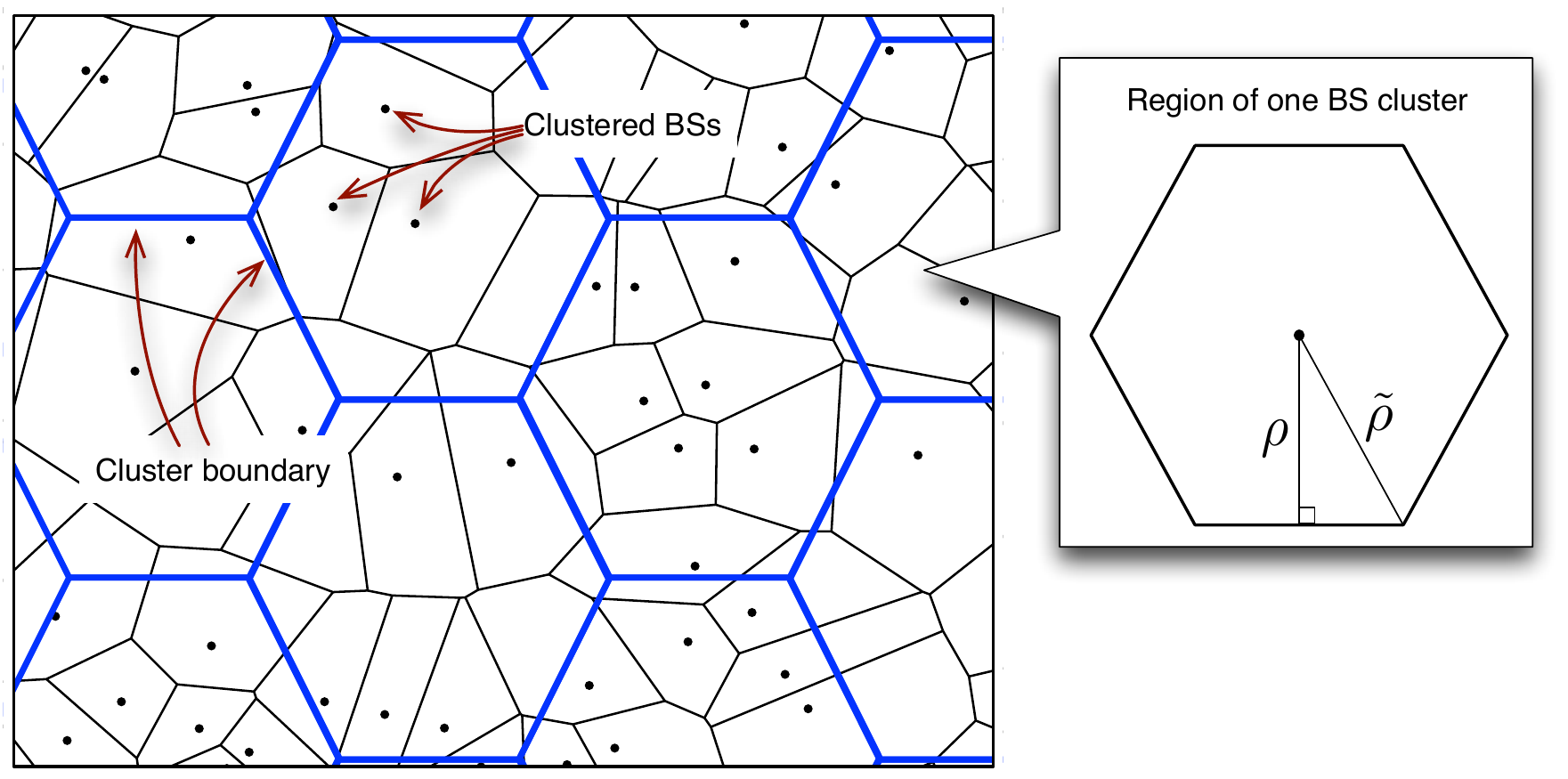}
\caption{(Left) The topology of the cellular network with Poisson distributed BSs clustered  using  a hexagonal lattice. The cells are drawn using thin lines and the cluster regions  thick lines; BSs are marked using black dots. (Right) A hexagonal cluster region where $\rho$ and $\tilde{\rho}$ denote the distances from  the cluster center to an edge and a vertex, respectively, and $\rho = \frac{\sqrt{3}}{2}\tilde{\rho}$. The cluster area is  $2\sqrt{3}\rho^2$ and hence the expected BS-cluster size is $\ell = 2\sqrt{3}\rho^2\lambda$.}
\label{Fig:NetTopology}
\end{center}
\end{figure*}

The BSs are modeled as  a homogeneous PPP $\Phi=\{Y\}$ in the horizontal plane $\mathds{R}^2$ with density $\lambda$ where  $Y\in\mathds{R}^2$ is  the coordinates of the corresponding BS.  The  mobiles form a homogeneous  point process  independent with $\Phi$. By assigning mobiles to their nearest BSs,  the horizontal plane  is partitioned into Voronoi cells as illustrated in Fig.~\ref{Fig:NetTopology}. It is assumed that the mobile density is much larger than the BS density such that each  cell contains at least one mobile almost surely. 
Each BS $Y$ serves a single mobile at a time, denoted as $u(Y)$,  selected from mobiles in the corresponding cell by uniform sampling. 
Consequently, the distance between an arbitrary BS $Y\in\Phi$ to the intended mobile, denoted as $L_Y$, has the following  distribution function \cite{Andrews:TractableApproachCoverageCellular:2010}:
\begin{equation}\label{Eq:L:Dist}
\Pr( L_Y > x) = e^{-\pi \lambda x^2}, \qquad x \geq 0. 
\end{equation}

BSs are clustered using a hexagonal lattice $\Omega = \{T\}$ where $T\in \mathds{R}^2$ denotes  the coordinates of a lattice point. 
Using  the lattice points as \emph{cluster centers}, the horizontal plane is partitioned into hexagonal \emph{cluster regions} as illustrated in Fig.~\ref{Fig:NetTopology}. Let $\mathcal{C}(T, r)$ denote a hexagon centered at $T\in\mathds{R}^2$ and having the distance $r$ from $T$ to the boundary. Thus the cluster region centered at   $T \in \Omega$ can be represented by  $\mathcal{C}(T , \rho)$  where $\rho$ is specified  in  Fig.~\ref{Fig:NetTopology}. Note that $\rho$ determines the density of the lattice $\Omega$. The area of $\mathcal{C}(T, \rho)$ is  $2\sqrt{3}\rho^2$ and hence the expected BS-cluster size is $\ell = 2\sqrt{3}\rho^2\lambda$.   Let $Y^*$ denote a typical point in $\Phi$, called the \emph{typical BS},  and the mobile served by $Y^*$ is called the \emph{typical mobile} and represented by  $U^*$. Moreover, define the typical cluster center $T^* \in \Omega$ as one such that  $\mathcal{C}(T^*, \rho)$ contains $Y^*$. The cluster of BSs lying in $\mathcal{C}(T^*, \rho)$, namely $\Phi\cap\mathcal{C}(T^*, \rho)$,  is called the \emph{typical BS cluster}; the associated cluster of mobiles is represented by $\mathcal{U}^* = \l\{u(Y)\mid Y \in \Phi\cap\mathcal{C}(T, \rho)\r\}$. 

\subsection{Multi-Cell Transmission}\label{Section:MultiCell:TX} The cooperation in a BS cluster is realized using a practical interference-coordination approach  that requires no inter-cell data exchange \cite{Gesbert:MultiCellMIMOCooperativeNetworks:2010}. Consider the typical  BS cluster $\Phi\cap\mathcal{C}(T^*, \rho)$  and the affiliated cluster of mobiles $\mathcal{U}^*$. Assume that each BS employs  $Q$ antennas and mobiles have  single-antennas.  Let $M$ denote the number of BSs and hence  $M= |\Phi\cap \mathcal{C}(T^*, \rho)|$ is a Poisson random variable (rv) with mean $\ell$. It is assumed that  $Q \geq M$ so that each BS has sufficient antennas for suppressing interference to mobiles served by other cooperating BSs. As a result, $Q$ is a rv and varies over different clusters. { The analysis in the sequel focuses on the regime of a large average cluster size ($\ell\rightarrow\infty$) corresponding to the regime of large-scale antenna arrays ($\E[Q] \rightarrow \infty$). With  expected deployment of large-scale arrays in future wireless networks \cite{RusekLarssonMarz:ScaleUpMIMO:2012}, such an assumption may be viable. Furthermore, the analytical results will be shown to also be  accurate for moderate numbers of antennas. For instance, it will be  observed subsequently from  simulation results (see Fig.~\ref{Fig:Pout}) that for sparse scattering the derived asymptotic bounds on the OPE are tight for $\ell$  smaller than $6$ and   $\E[Q]$ equal to $\ell$ plus several more antennas to achieve moderate array gain.}
Let $h_{UY}^{[k]}\in \mathds{C}$ represent the coefficient of the scalar channel from the $k$-th antenna at $Y$ to $U$ and define the channel vector $\bh_{UY}=\big[h_{UY}^{[1]}, h_{UY}^{[2]}, \cdots, h_{UY}^{[Q]}\big]^T$ for given $Q$. Moreover, let $\bff_Y\in \mathds{C}^Q$ with $\|\bff_Y\| = 1$ denote the unitary transmit beamformer used at  $Y$.  The interference avoidance at $Y^*$ is achieved by  choosing  $\bff_Y$ to be orthogonal to the $(M-1)$ interference channels and   the remaining $N = Q-M + 1$ degrees of freedom (DoF), called the \emph{diversity order},  are applied  to attain diversity gain \cite{Jindal:RethinkMIMONetwork:LinearThroughput:2008}. { It is assumed that $N \geq \nu $ with $\nu > 1$ being  the minimum diversity order over different cells, where the constraint $\nu > 1$ ensures finite average transmission power under channel inversion for the case of rich scattering.} Assuming perfect CSI at BSs, their beamformers are designed using the zero-forcing criterion as follows. 
\begin{definition}[Interference coordination]\label{Def:IC:Beam}\emph{Conditioned on $Q = m$, the beamformer  $\bff_{Y^*}$ used at the typical  BS $Y^*$ solves:
\begin{equation}\label{Eq:ZFBeam}
\begin{aligned}
\text{maximize:} \quad & |\bff^\dagger \bh_{U^* Y^*}| \\
\text{subject to:} \quad & \bff^\dagger \bh_{UY^*}=0 \ \forall \ U\in \mathcal{U}^*\backslash\{U^*\}\\
& \bff\in \mathds{C}^m, \|\bff\| =1. 
\end{aligned}
\end{equation}}
\end{definition}
\noindent 

{ This algorithm is also considered in \cite{Zhang:AdaptSpatialICMulticell:2010} for mitigating inter-cell interference  in a two-cell network.} Note that the computation of $\bff_{Y^*}$ requires $Y^*$ to have CSI of both the data channel and the $(M-1)$  channels from $Y^*$ to mobiles served by other cooperating BSs, which can be acquired by CSI feedback \cite{Love:OverviewLimitFbWirelssComm:2008}. Given that the network is interference limited, with  the beamformer designed as  in Definition~\ref{Def:IC:Beam}, the signal $y$ received at $U^*$ is given as
\begin{equation}\label{Eq:InOut:IC}
y \!=\! \sqrt{P_{Y^*}} \bff_{Y^*}^\dagger \bh_{U^* Y^*} x_{U^*} +\!\!\!\! \sum_{Y\in \Phi\cap\bar{\mathcal{C}}(T^*, \rho)}\!\!\!\!\sqrt{P_Y}\bff_Y^\dagger \bh_{U^* Y}x_{u(Y)}
\end{equation}
where $P_Y$ denotes the transmission power of BS $Y$ and $x_U$ is a data symbol with unit variance and intended for $U$. 
Let $S$ and $I$ represent the signal and interference powers measured at $U^*$, respectively. It follows from \eqref{Eq:InOut:IC} that 
\begin{equation}
S = P_{Y^*} \bigl|\bff_{Y^*}^\dagger \bh_{U^* Y^*}\bigr|^2 \quad \text{and}\quad I =  \!\!\!\!\!\!\sum_{Y\in \Phi\cap\bar{\mathcal{C}}(T^*, \rho)}P_Y\bigl|\bff_Y^\dagger \bh_{U^* Y}\bigr|^2. \label{Eq:RxPwr:IC}
\end{equation}

Besides mitigating interference using MCC,  channel inversion is  applied at BSs to cope with data-link fading. The transmission power $P_Y$  of BS $Y$   is chosen such that the signal power received  by the intended mobile is a constant $\omega > 0$. Consequently, $S = \omega$ and 
\begin{equation}\label{Eq:PwrCtrl}
P_{Y^*} = \frac{\omega}{|\bff_{Y^*}^\dagger \bh_{U^* Y^*}|^2}
\end{equation}
where $\omega$ satisfies the average power constraint $\E[P_{Y^*}]\leq \bar{P}$ with $\bar{P} >0$ and hence is given as 
\begin{equation}\label{Eq:PwrCtrl:RxPwr}
\omega = \frac{\bar{P}}{\E\bigl[|\bff_{Y^*}^\dagger \bh_{U^* Y^*}|^{-2}\bigr]}. 
\end{equation}
It is found in this research that channel inversion increases OPE (reduces outage probability) compared with fixed-power transmission. { The reason is that fixed-power transmission causes  fluctuation in received signal power that increases outage probability but  can be removed by channel inversion.}  The analysis for the scenario  of fixed-power transmission is omitted to keep  the exposition precise. 

\subsection{Channel Models} \label{Section:Channel}
The scattering environment affects the interference distribution and hence the OPE. For this reason, both sparse and rich scattering are considered in the OPE analysis and their models are described as follows.

\subsubsection{Sparse Scattering} 
In an environment with sparse scatterers, there usually exists a line-of-sight path between a transmitter and a  receiver and fading is negligible compared with this direct path.
 Using beamforming  in Definition~\ref{Def:IC:Beam},  each multi-antenna BS forms a physical beam such that the main lobe is steered  towards the intended mobile, nulls towards mobiles served by cooperating BSs, and side-lobes towards others \cite{VanVeen:Beamforming:1988}.  This can be modeled such that the  interference power $I$  in \eqref{Eq:RxPwr:IC} and transmission power $P_{Y^*}$ in  \eqref{Eq:PwrCtrl} are given as 
\begin{align}
I &=  \sum_{Y\in \Phi\cap\bar{\mathcal{C}}(T^*, \rho)}P_Y G_{U^* Y} |Y - U^*|^{-\alpha} \label{Eq:IntPwr:Sparse}, \\
P_{Y^*} &= \omega L_{Y^*}^{\alpha}W_{Y^*}^{-1} \label{Eq:SigPwr:Sparse}
\end{align}
where the path-loss exponent $\alpha > 2$, $W_Y$ is the main-lobe response of beamforming  at $Y$, and $G_{UY}$ is its side-lobe response in the direction from $Y$ to $U$. In practice, the values of $W_{Y^*}$ and $G_{U^* Y}$ depend on the size  and configuration  of BS antenna arrays as well as transmission directions \cite{VanVeen:Beamforming:1988}. They  are modeled as random variables (rvs)  with the following properties. 

\begin{assumption}[Sparse-scattering model]  \label{AS:IC:Sparse}\emph{
The rv  $W_{Y}$  has  bounded support $[\delta, \delta']$ with  $ \delta'\geq \delta > 0$.   For  $U$ and $Y$ associated with different BS clusters, the rv $G_{UY}$   has   bounded compact support $[0,  \gamma]$ with $\gamma > 0$. The set of rvs $\{G_{UY}\mid U\in\mathcal{U}^*, Y\in \Phi\cap\bar{\mathcal{C}}(T^*, \rho)\}$ are independent and identically distributed (i.i.d.).}
\end{assumption}

For clarification, the equality $\gamma=\delta'$ holds in theory since it is possible for a transmitter to direct a beam towards both an intended and an unintended receivers if they lie in the same direction. Nevertheless, given sufficiently sharp beams and randomly located nodes, such an event occurs with negligible probability and hence it can be assumed that $\gamma \ll \delta, \delta'$. This assumption, however, is not required for the current analysis.

\subsubsection{Rich Scattering} 
The channel is assumed to be frequency non-selective and follows  independent block fading.  Rich scattering is modeled by i.i.d. Rayleigh fading as follows.

\begin{assumption}[Rich-scattering model]\label{AS:Rich}\emph{An arbitrary channel  coefficient $h_{UY}^{[k]}$  is given as $h_{UY}^{[k]} = B^{[k]}_{UY}|U-Y|^{-\alpha}$ where $B_{UY}^{[k]}$ is a   $\mathcal{CN}(0, 1)$ rv. Any two rvs  $B^{[k]}_{UY}$ and $B^{[k']}_{U'Y'}$ with $(k, U, Y) \neq (k', U', Y')$ are independent. } 
\end{assumption}

It follows from Assumption~\ref{AS:Rich} that an arbitrary channel  vector $\bh_{UY}$ can be written as $\bh_{UY} = \bq_{UY}|U-Y|^{-\alpha}$ where $\bq_{UY}$ is a $Q\times 1$ random vector comprising i.i.d. $\mathcal{CN}(0, 1)$ elements. 
Moreover, the sequence $\{\bq_{UY}\}$ is i.i.d. 
The signal and interference powers measured at  $U^*$  are given by \eqref{Eq:IntPwr:Sparse} and \eqref{Eq:SigPwr:Sparse} but with the parameters $W_{Y^*}$ and $G_{U^* X}$ re-defined as 
$W_{Y^*} = |\bff_{Y^*}^\dagger \bq_{U^* Y^* }|^2$ and $G_{U^* Y}= |\bff_Y^\dagger \bq_{U^* Y}|^2$.
 The lemma below follows from    \cite[Lemma~$1$]{Jindal:RethinkMIMONetwork:LinearThroughput:2008} that studies zero-forcing beamforming (see Definition~\ref{Def:IC:Beam}) for  mobile ad hoc networks. 
\begin{lemma}[\cite{Jindal:RethinkMIMONetwork:LinearThroughput:2008}]\label{Lem:Jindal}\emph{For rich scattering and conditioned on $N=n$, $W_{Y^* }$  is a chi-square rv with $2n$ DoF and $\{G_{U^* Y}\mid Y\in \Phi\cap\bar{\mathcal{C}}(T^*, \rho)\}$ are i.i.d. exponential rvs with unit mean. }
\end{lemma}

\section{OPE with Sparse Scattering}\label{Section:OPE:Sparse}
In this section, the OPE is  analyzed for the environment of sparse scattering. Specifically, the OPE is characterized for a cluster-center mobile  and for a typical mobile separately. The results show that mobiles near cluster edges limit  network coverage.

\subsection{OPE for Cluster-Center Mobiles} 
Consider a mobile located at the typical  cluster center $T^* $ that is farthest from  the \emph{interference zone} among all mobiles and hence has the smallest outage probability, where an interference zone for a mobile refers to  a region in the horizontal plane comprising  interfering BSs.  The OPE for a  cluster-center mobile, denoted as $\phi^{cc}$,  can be written by modifying \eqref{Eq:OPE:Def} to account for the constraint  $U^* = T^*$: 
\begin{equation}\label{Eq:CCOPE}
\phi^{cc}(\ell) =-\log \Pr(I^{cc} > \theta^{-1}\omega\mid U^* = T^*) 
\end{equation}
where $I^{cc}$ represents the interference power measured at $T^*$. 
Asymptotic bounds on $\phi^{cc}$ for large $\ell$ are derived in the  sub-sections and then combined to give the main result of this section. 

\begin{figure*}
\begin{center}
\subfigure[Cluster-center mobile]{\includegraphics[width=8cm]{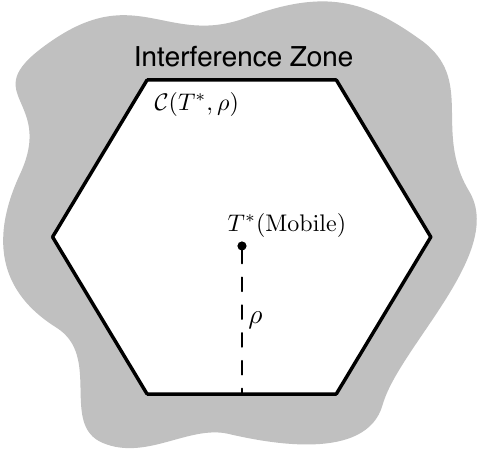}}\qquad \qquad
\subfigure[Typical mobile]{\includegraphics[width=8cm]{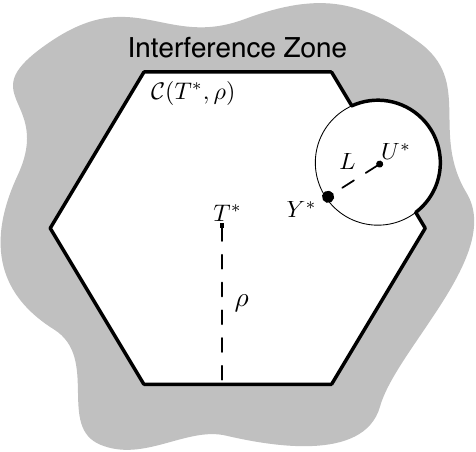}}
\caption{The geometry of the interference zones (shaded regions) for (a) a cluster-center mobile and (b) a typical mobile,   where the interferers are Poisson distributed  with density $\lambda$. The symbols $T^*$, $\mathcal{C}(T^*, \rho)$, $Y^*$, and  $U^*$ represent the typical cluster center, cluster region, BS and mobile, respectively, and $L=|Y^* - U^*|$. }
\label{Fig:Interferers}
\end{center}
\end{figure*}

\subsubsection{Asymptotic Lower Bound on the OPE}\label{Section:ALB:CCOPE:Sparse} First, a lower bound on $\phi^{cc}$ is obtained as follows.  Slightly abusing notation, let $T^*$ also represent the typical cluster-center mobile.  As illustrated in Fig.~\ref{Fig:Interferers}(a), $\bar{C}(T^*, \rho)$  is  the complete  interference zone for $T^*$. 
Therefore, $I^{cc}$  can be obtained by modifying  \eqref{Eq:IntPwr:Sparse} as 
\begin{equation}\label{Eq:iPower:IC:Rich}
I^{cc} = \sum_{Y\in\Phi\cap\bar{\mathcal{C}}(T^*, \rho)}P_Y G_{T^* Y} |Y - T^*|^{-\alpha}  
\end{equation}
which is a \emph{power-law-shot-noise} process \cite{Lowen:PowerLawShotNoise:1990}. 
It can be observed from \eqref{Eq:CCOPE} that the  OPE is determined by  the tail probability of $I^{cc}$ that, however, has no  closed-form expression \cite{Lowen:PowerLawShotNoise:1990}. For the current analysis, it suffices by deriving  an upper bound  on $I^{cc}$. This relies on decomposing  $I^{cc}$ into a series of compound Poisson rvs  inspired by the  approach in \cite{Ganesh:LargeDeviationInterferenceWirelessNet}. To this end,   the interference zone  $\bar{\mathcal{C}}(T^*, \rho)$ is partitioned into a sequence of disjoint hexagonal rings $\{\mathcal{A}_n\}_{n=1}^\infty$  with $\mathcal{A}_n = \mathcal{C}(T^*, \sqrt{n+1} \rho) \backslash \mathcal{C}(T^*, \sqrt{n} \rho)$. Note that $\{\mathcal{A}_n\}$ have the same  area as  $\mathcal{C}(T^*, \rho)$. The interference  power measured at $T^*$ due to interferers lying  in  $\mathcal{A}_n$ is represented by
\begin{equation}\label{Eq:In}
I_n^{cc}  = \sum_{Y\in \Phi \cap \mathcal{A}_n} P_Y G_{T^* Y} |Y - T^*|^{-\alpha}.
\end{equation}
Therefore, $I^{cc}$ in  \eqref{Eq:iPower:IC:Rich} can be decomposed as   $I^{cc}  = \sum_{n=1}^{\infty}I_n^{cc}$. To facilitate analysis, define a compound Poisson rv $Z_n$ as 
\begin{equation}\label{Eq:Zn:Def}
Z_n = \sum_{Y\in \Phi \cap \mathcal{A}_n} P_Y G_{U^* Y}
\end{equation}
where $\{P_Y G_{U^* Y}\}$ are i.i.d. and the number of terms in the summation, namely $| \Phi \cap \mathcal{A}_n|$, is a Poisson rv with mean $\ell$. Note that the distribution of $Z_n$ is independent of $U^*$. Based on the geometry of  $\mathcal{A}_n$, it can be obtained from \eqref{Eq:In} that 
$I_n^{cc}  \leq (\sqrt{n}\rho )^{-\alpha} Z_n$.
Since $\ell = 2\sqrt{3}\rho^2\lambda$, it follows that 
\begin{equation}\label{Eq:IPwr:UB}
I^{cc} \leq \l(\frac{2\sqrt{3}\lambda}{\ell}\r)^\frac{\alpha}{2}\sum_{n=1}^\infty n^{-\frac{\alpha}{2}} Z_n. 
\end{equation}
Combining \eqref{Eq:CCOPE} and \eqref{Eq:IPwr:UB}  yields a  lower bound on $\phi^{cc}$:
\begin{equation}\label{Eq:CCOPE:LB:Sparse}
\phi^{cc}(\ell) \geq  -\log \Pr\l(\sum_{n=1}^\infty n^{-\frac{\alpha}{2}} Z_n > \frac{\omega}{\theta(2\sqrt{3}\lambda)^\frac{\alpha}{2}}\ell^{\frac{\alpha}{2}}\r). 
\end{equation}

Next, an asymptotic lower bound  on $\phi^{cc}(\ell)$  as $\ell\rightarrow\infty$ can be derived by analyzing  the large deviation    of the summation  in \eqref{Eq:CCOPE:LB:Sparse}  as follows. As $Z_n$ is a sum over  the i.i.d. sequence   $\{P_Y G_{T^* Y}\mid Y \in \Phi\cap \mathcal{A}_n\}$, it is necessary to characterize  the large deviation  of $P_Y G_{U^* Y}$ as follows. 

\begin{lemma}\label{Lem:TXPwr:Tail} \emph{For sparse scattering and an arbitrary  BS $Y\in \Phi\cap\bar{\mathcal{C}}(T^*, \rho)$,  $\E[P_Y G_{U^* Y}]$ is finite and
\begin{equation}\label{Eq:PG:Tail}
\!\!-\log \Pr( P_Y G_{U^* Y} > x) \sim \pi\lambda\l(\frac{\delta}{\gamma\omega}\r)^{\frac{2}{\alpha}} x^{\frac{2}{\alpha}}, \quad x \rightarrow\infty.
\end{equation} 
}
\end{lemma}
The proof of Lemma~\ref{Lem:TXPwr:Tail} is given in Appendix~\ref{App:TXPwr:Tail}.  Analyzing the large deviation of $Z_n$ also requires the following result from 
\cite[Proposition~$7.1$]{Mikosch:LargeDevHeavyTailedInsurance:1998}. 
\begin{lemma}[\cite{Mikosch:LargeDevHeavyTailedInsurance:1998}]\label{Lem:CompPoisson} \emph{Consider a compound Poisson rv $Z_0 = \sum_{m=1}^{F} H_m$
where $F$ follows the Poisson distribution  and  $\{H_m\}$ are i.i.d. rvs independent of  $F$. 
If the distribution of $H_m$ is  either   $\mathsf{RV}(\tau)$ with $\tau > 0$  or $\mathsf{WE}(\tau)$ with $0 < \tau < 0.5$,
\begin{equation}\nn
\Pr(Z_0 - \E[Z_0] > x)\sim \E[F]\Pr(H_1 > x), \qquad \E[F] \rightarrow \infty
\end{equation}
if $x >  a \E[Z_0]$ for all $a > 0$, where $\E[Z_0] = \E[F]\E[H_1]$. }
\end{lemma}
Since $\{P_Y G_{U^* Y}\}\in \mathsf{WE}(\tau)$ with $0 < \tau < 0.5$ from Lemma~\ref{Lem:TXPwr:Tail}, using the definition of $Z_n$ in \eqref{Eq:Zn:Def} and applying Lemma~\ref{Lem:CompPoisson} lead to the following result that is  proved in Appendix~\ref{App:LargeDev:Z}.  
\begin{lemma}\label{Lem:LargeDev:Z}\emph{Given $x > 0$, if $\alpha > 4$, 
\begin{equation}\label{Eq:LogZn:LB:a}
-\log \Pr\l(Z_n > \ell^{\frac{\alpha}{2}} x\r) \sim \pi\lambda\l(\frac{ \delta}{\gamma\omega}\r)^{\frac{2}{\alpha}} \ell x^{\frac{2}{\alpha}}, \qquad \ell \rightarrow\infty, 
\end{equation}
 and if $2 < \alpha \leq 4$, 
\begin{equation}\label{Eq:LogZn:LB:b}
-\log \Pr\l(Z_n > \ell^{\frac{\alpha}{2}} x\r) 
\succeq \pi\lambda\l(\frac{ \delta}{\gamma\omega}\r)^{\frac{2}{\alpha}} \ell^{\frac{\alpha}{4}}\sqrt{x}, \qquad \ell \rightarrow\infty. 
\end{equation} 
where $n=1, 2, \cdots$.  
}
\end{lemma}

Given Lemma~\ref{Lem:LargeDev:Z},   the application of the \emph{contraction principle} from large-deviation theory  (see e.g., \cite[Theorem~$4.2.1$]{DemboBoo:LargeDeviation}) yields \footnote{The procedure is similar to that for obtaining \eqref{Eq:Lem:Zn:1} in Appendix~\ref{App:LargeDev:Z}. } 
\begin{equation}\label{Eq:Zn:Z}
\!\!-\log \Pr\l(\sum_{n=1}^\infty n^{-\frac{\alpha}{2}}Z_n > \ell^{\frac{\alpha}{2}} x\r)  \!\sim\! -\log \Pr\l(Z_n \!>\! \ell^{\frac{\alpha}{2}} x\r)
\end{equation}
as  $\ell \rightarrow\infty$. 
Combining  \eqref{Eq:CCOPE:LB:Sparse}, \eqref{Eq:Zn:Z} and Lemma~\ref{Lem:LargeDev:Z} leads to an asymptotic lower bound on $\phi^{cc}$ as shown below.  

\begin{lemma}\label{Lem:CCOPE:LB:Sparse}\emph{As $\ell\rightarrow\infty$, the OPE for a  cluster-center mobile satisfies 
\begin{equation}\label{Eq:OPE:ALB:Sparse}
\phi^{cc}(\ell) \succeq \l\{
\begin{aligned}
&c_1 \ell, && 
\alpha > 4\\
&c_2 \ell^{\frac{\alpha}{4}}, && 2 < \alpha \leq 4 
\end{aligned}
\r.
\end{equation}
where the constants $c_1$ and $c_2$ are defined as
\[
c_1 = \frac{\pi}{2\sqrt{3}}\l(\frac{ \delta}{\theta\gamma}\r)^{\frac{2}{\alpha}},\qquad 
c_2 = \frac{\pi\lambda^{1-\frac{\alpha}{4}}\delta^{\frac{2}{\alpha}}}{\omega^{\frac{4-\alpha}{2\alpha}}\sqrt{\theta}(2\sqrt{3})^{\frac{\alpha}{4}}\gamma^{\frac{2}{\alpha}}}.  
\]
}
\end{lemma}

\subsubsection{Asymptotic Upper Bound on the OPE} The OPE $\phi^{cc}$ can be upper bounded by considering only the  interferers for $T^*$ from a subset of the interference zone $\bar{C}(T^*, \rho)$. For this purpose, define a ``narrow" hexagonal  ring 
\begin{equation}
\mathcal{A}_\epsilon = \mathcal{C}(T^*, \sqrt{1+\epsilon} \rho) \backslash \mathcal{C}(T^*,  \rho)
\end{equation}
with $\epsilon > 0$ and 
\begin{equation}\label{Eq:Z:Eps}
Z_{\epsilon} = \sum_{Y \in \Phi \cap \mathcal{A}_\epsilon} P_Y G_{T^* Y}. 
\end{equation}
Note that $Z_{\epsilon}$ is a compound Poisson rv where the Poisson distribution has mean $\epsilon \ell$.
Since $|Y - T^* | \leq (1+\epsilon) \tilde{\rho}$ for all $Y \in \Phi \cap \mathcal{A}_\epsilon$  and $\mathcal{A}_\epsilon \in \bar{\mathcal{C}}(T^*,  \rho)$, $I^{cc}$  in \eqref{Eq:iPower:IC:Rich} is lower bounded as  
\begin{equation}
I^{cc} \geq [(1+\epsilon)\tilde{\rho}]^{-\alpha}Z_{\epsilon}. \label{Eq:I:LB:Rich:NoPC}
\end{equation}
 By combining  \eqref{Eq:CCOPE} and \eqref{Eq:I:LB:Rich:NoPC} and using $\tilde{\rho} = \frac{2}{\sqrt{3}}\rho$,  the OPE $\phi^{cc}$ can be upper bounded as
\begin{equation}
\phi^{cc}(\ell)  \leq -\log \Pr\l(\l(\frac{2}{\sqrt{3}}(1+\epsilon)\rho\r)^{-\alpha}Z_{\epsilon} > \theta^{-1}\omega\r). \label{Eq:CCOPE:UB:Sparse:a}
\end{equation}
Analyzing the scaling of the  right-hand side of  \eqref{Eq:CCOPE:UB:Sparse:a}  as $\ell\rightarrow\infty$ leads to an asymptotic upper bound on $\phi^{cc}$ as shown in Lemma~\ref{Lem:CCOPE:UB:Sparse} that is proved in Appendix~\ref{App:CCOPE:UB:Sparse}. 
\begin{lemma}\label{Lem:CCOPE:UB:Sparse}\emph{For sparse scattering and as $\ell\rightarrow\infty$, the OPE for a cluster-center mobile satisfies  
\begin{equation}\label{Eq:OPE:UB:Sparse}
\phi^{cc}(\ell) \preceq \frac{4c_1}{3} \ell
\end{equation}
where the constant $c_1$ is as defined in Lemma~\ref{Lem:CCOPE:LB:Sparse}. 
}
\end{lemma}

\subsubsection{Main Result and Remarks} Combining Lemma~\ref{Lem:CCOPE:LB:Sparse} and \ref{Lem:CCOPE:UB:Sparse} leads to the following theorem. 
\begin{theorem} \label{Theo:CCOPE:Sparse} 
\emph{For    sparse scattering and as  $\ell\rightarrow\infty$, the OPE for  a cluster-center mobile satisfies 
\begin{enumerate}
\item for $\alpha > 4$, 
\begin{equation}\label{Eq:CCPout:Sparse:PC:a} 
\eqbox{
c_1 \ell \preceq\phi^{cc}(\ell) \preceq  \frac{4c_1}{3} \ell, 
}
\end{equation}
\item and for   $2 < \alpha \leq 4$,  
\begin{equation}\label{Eq:CCPout:Sparse:PC:b} \eqbox{
c_2 \ell^{\frac{\alpha}{4}}  \preceq \phi^{cc}(\ell) \preceq \frac{4c_1}{3} \ell,
}
\end{equation} 
where $c_1$ and $c_2$ are as defined in Lemma~\ref{Lem:CCOPE:LB:Sparse}. 
\end{enumerate}}
\end{theorem}

Several remarks are in order. 

\begin{enumerate}
\item Theorem~\ref{Theo:CCOPE:Sparse} shows that  $\phi^{cc}(\ell)$ scales \emph{linearly} with  increasing $\ell$ for a large path-loss exponent ($\alpha > 4$) and at least \emph{sub-linearly} for a moderate-to-small  exponent $(2 < \alpha \leq 4)$. These results  suggest that as $\ell\rightarrow\infty$, $\Pout$ diminishes exponentially and at least sub-exponentially  for  $\alpha > 4$ and $ 2 < \alpha \leq 4$, respectively. The scaling of  $\Pout$ depends on  $\alpha$ because it determines the level of spatial separation. Note that for $\alpha > 4$, the asymptotic bounds on $\phi^{cc}(\ell)$ have a ratio of $4/3$ [see \eqref{Eq:CCPout:Sparse:PC:a}] and hence are tight.   Mathematically, the tightness of the bounds is due to the  product rv $P_YG_{T^*Y}$ in the expression for $I^{cc}$ in \eqref{Eq:iPower:IC:Rich} having a distribution with  a sufficiently heavy right tail, allowing accurate characterization  of  the asymptotic tail probability of $I^{cc}$. However, as $\alpha$ decreases, {the tail probability of $P_YG_{T^*Y}$ reduces.} This results in that the ratio of the asymptotic bounds on $\phi^{cc}(\ell)$ in \eqref{Eq:CCPout:Sparse:PC:b} diverges  as $\ell$ increases.

\item It can be observed from  Theorem~\ref{Theo:CCOPE:Sparse} and the definitions of $c_1$ and $c_2$ that larger $\phi^{cc}(\ell)$ results from increasing the ratio $\delta/\gamma$, namely the minimum ratio between the  magnitudes  of beam  main-lobes and side-lobes. In other words, as $\ell\rightarrow\infty$, the outage probability diminishes  faster for sharper beams, agreeing  with intuition.

\item  Theorem~\ref{Theo:CCOPE:Sparse} suggests that for fixed outage probability,  the outage threshold $\theta$  should be  proportional to  $\ell^{\frac{\alpha}{2}}$. Correspondingly, the throughput of a cluster-center mobile, defined  as $R=\log(1+\theta)$,   grows with increasing $\ell$ as 
\[
R \sim \frac{\alpha}{2}\log \ell, \qquad \ell\rightarrow\infty.
\] 
Note that  $R$ scales  linearly with $\alpha$ because large $\alpha$ corresponds to more severe attenuation of  inter-cluster interference. 
\end{enumerate}

\subsection{OPE for Typical Mobiles}
Consider the typical mobile $U^*$ and the corresponding OPE $\phi$ as given in \eqref{Eq:OPE:Def}. The asymptotic bounds on $\phi$  are derived in the following subsections. 

\begin{figure*}
\begin{center}
\includegraphics[width=10cm]{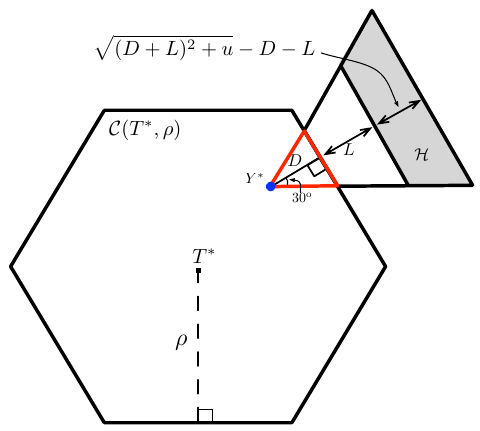}
\caption{Geometric definition  of the set $\mathcal{H}\subset\mathds{R}^2$ used in the proof of Lemma~\ref{Lem:OPE:UB:Sparse}. }
\label{Fig:Hex_Proof_2}
\end{center}
\end{figure*}

\subsubsection{Asymptotic Lower Bound on the OPE}
First, a lower bound on $\phi$ is obtained as follows. 
For ease of notation, the distance $L_{Y^*}$ between the typical mobile and BS  is re-denoted as $L$. 
As illustrated in Fig.~\ref{Fig:Interferers}(b),  the interfering BSs for $U^* $  are Poisson distributed in   the region $\Lambda = \bar{\mathcal{C}}(T^*, \rho) \cap \bar{\mathcal{O}}(U^* , L)$ where $\mathcal{O}(A, r)$ represents a disk centered at $A\in\mathds{R}^2$ and with the radius $ r\geq 0$, namely that  $\mathcal{O}(A, r) = \{X\in\mathds{R}^2\mid |X - A| \leq r\}$. Note that $\bar{\mathcal{O}}(U^* , L)$ encloses  the complete  interference zone for $U^* $ 
due to the fact that any interfering BS for $U^* $ is farther than the serving BS $Y^* $ at a distance of $L$ from $U^*$. As a result, the interference power for $U^* $ can be written as 
\begin{align}
I &= \sum_{Y\in \Phi\cap\Lambda} P_Y G_{U^*Y} |Y - U^* |^{-\alpha}\nn\\
&\leq \sum_{Y\in \Phi\cap\Lambda} P_Y G_{U^*Y} |Y - U^* |^{-\alpha} + \nn\\
&  \sum_{Y\in \Phi\cap\bar{\mathcal{C}}(T^*, \rho)\cap\mathcal{O}(T^*, \rho)}\!\!\! P_Y G_{U^*Y} \l[\max(|Y - U^* |, L)\r]^{-\alpha}\!\!.\label{PF:T2:Eq1:b}
\end{align}
Using the facts that  $\bar{\mathcal{C}}(T^*, \rho)=\Lambda \cup\l[\bar{\mathcal{C}}(T^*, \rho)\cap\mathcal{O}(T^*, \rho)\r]$ and $|Y - U^* | = \max(|Y - U^* |, L)$ if $Y\in \Phi\cap\Lambda$ [see Fig.~\ref{Fig:Interferers}(b)],  it follows from \eqref{PF:T2:Eq1:b}: 
\begin{align}
I &\leq \sum_{Y\in \Phi\cap\bar{\mathcal{C}}(T^*, \rho)} P_Y G_{U^*Y} \l[\max(|Y - U^* |, L)\r]^{-\alpha}\label{PF:T2:Eq1:a}\\
&\leq \sum_{n=1}^\infty \sum_{Y\in \mathcal{A}_n} P_Y G_{U^*Y} \l[\max(|Y - Y^* |-L, L)\r]^{-\alpha}
\label{PF:T2:Eq1}
\end{align}
\eqref{PF:T2:Eq1} is obtained based on  the triangular inequality $|Y-U^* | \geq |Y- Y^* | - |Y^*  - U^* |$  and  $\bar{\mathcal{C}}(T^*, \rho) = \cup_{n}\mathcal{A}_n$ with $\{\mathcal{A}_n\}$ being the hexagonal rings defined in Section~\ref{Section:ALB:CCOPE:Sparse}. Let $D$ denote the  distance from  $Y^* $ to  the boundary  of $\mathcal{C}(T^*, \rho)$:  $D = \min_{X\in\bar{\mathcal{C}}(T^*,  \rho)} |Y^*-X |$. By the stationarity of the mobile and BS processes,   $Y^* $ is uniformly distributed in $\mathcal{C}(T^*, \rho)$, resulting in the following  distribution of $D$:
\begin{equation}
\Pr(D \leq x) = 1 - \l(1 - \frac{x}{\rho}\r)^2, \qquad  0 \leq x \leq \rho. \label{Eq:Dist:D} 
\end{equation}
Since the shortest distance between $Y^* $ and a point in  $\mathcal{A}_n$ is $\sqrt{n}\rho - \rho + D$, it follows from \eqref{PF:T2:Eq1} that
\begin{equation}
I \leq \sum_{n=1}^\infty \l[\max(\sqrt{n}\rho - \rho + D-L, L)\r]^{-\alpha}Z_n \label{PF:T2:Eq2}
\end{equation}
where $Z_n$ is defined in \eqref{Eq:Zn:Def}.  From \eqref{Eq:OPE:Def} and \eqref{PF:T2:Eq2}, $\phi$ can be lower bounded as
\begin{equation}\label{Eq:OPE:LB:Sparse:a}
\begin{aligned}
\phi(\ell) \geq -\log\Pr\Bigg(\sum_{n=1}^\infty &\big(\max(\sqrt{n}\rho - \rho +\\
& D-L, L)\big)^{-\alpha}Z_n > \theta^{-1}\omega\Bigg). 
\end{aligned}
\end{equation}

Next, an asymptotic lower bound on $\phi(\ell)$ is derived by analyzing the scaling of the right-hand size of \eqref{Eq:OPE:LB:Sparse:a} as $\ell\rightarrow\infty$. For this purpose, 
it is shown in the following lemma that  $\phi(\ell)$ can be asymptotically upper bounded by an expression comprising a series of the i.i.d. compound Poisson rvs $\{Z_n\}$, which facilitates     a similar approach as used for obtaining Lemma~\ref{Lem:CCOPE:LB:Sparse}. 

\begin{lemma}\label{Lem:PoutUB:Sparse:a}\emph{For sparse scattering and as $\ell \rightarrow\infty$, the OPE for a typical mobile satisfies
\begin{align}
\phi(\ell) \succeq  \min&\Bigg(\max_{z > 0}\min\Bigg(-\log\Pr\l(\sum_{n=1}^\infty n^{-\frac{\alpha}{2}} Z_n >  \frac{\omega y}{2^{\alpha}\theta}\r), \nn\\
&-\log \Pr\l(D^{\alpha} \leq y\Bigg)\Bigg)\!,  -\log\Pr\l(L >  \frac{D}{2}\r)\r)\!. \nn
\end{align}
}
\end{lemma}
The proof of Lemma~\ref{Lem:PoutUB:Sparse:a} is provided in Appendix~\ref{App:PoutUB:Sparse:a}. By analyzing the scalings of the three terms in the lower bound on $\phi$,   an asymptotic lower bound on the OPE is obtained as follows. 

\begin{lemma}\label{Lem:OPE:LB:Sparse}\emph{For sparse scattering and as $\ell\rightarrow\infty$, the OPE for a typical mobile satisfies 
\begin{equation}
\phi(\ell) \succeq \frac{1}{2}\l(1 - \frac{2}{\alpha}\r)\log \ell. 
\end{equation}
}
\end{lemma}
The proof of Lemma~\ref{Lem:OPE:LB:Sparse} is  provided in Appendix~\ref{App:OPE:LB:Sparse}.

\subsubsection{Asymptotic Upper Bound on the OPE} The analytical technique for deriving an upper bound on  $\phi$ essentially considers only interference to $U^*$ from interferers lying in  a subset of the interference zone $\Lambda$ defined in the preceding section. Specifically, define  a region $\mathcal{H}\subset \Lambda$ (see Fig.~\ref{Fig:Hex_Proof_2}) as
\begin{align}
\mathcal{H} =\Big \{X \in \bar{\mathcal{C}}&(Y^*, D+L)\cap \mathcal{C}(Y^*, \sqrt{(D+L)^2 + u}) \mid \nn\\
&- \frac{\pi}{6} \leq \angle(X - Y^* )-\angle(J_{Y^*} - Y^* )  \leq \frac{\pi}{6}\Big\} \nn
\end{align}
where $u > 0$ and $J_{Y^*}$ is a point in  $\bar{\mathcal{C}}(T^*, \rho)$ such that $|J_{Y^*} - Y^* |= D$. Note that the hexagons in the definition of $\mathcal{H}$ are chosen such that the area of $\mathcal{H}$ is a constant $u/\sqrt{3}$.  Then the OPE $\phi$ in \eqref{Eq:OPE:Def} can be upper bounded as 
\begin{equation}
\begin{aligned}
\phi(\ell) \leq  -\log\Big(&\Pr\Big(\sum\nolimits_{Y\in\mathcal{H}} P_{Y}G_{U^*Y}|Y-U^*|^{-\alpha} > \\
&\theta^{-1}\omega \mid \Phi \cap \mathcal{H}\neq \emptyset\Big)\Pr(\Phi \cap \mathcal{H}\neq \emptyset)\Big).
\end{aligned}
\label{Eq:OPE:UB:Sparse:PF}
\end{equation}
Let $Y_0$ denote an arbitrary BS in $\mathcal{H}$ conditioned on $\Phi\cap\mathcal{H}\neq \emptyset$. Since $|Y_0 - U^* | \leq |Y_0 - Y^* | + L$ by the triangular inequality  and  
\begin{equation}
|Y_0 - Y^*|\leq \frac{\sqrt{3}}{2}\sqrt{(D+L)^2 + u}
\end{equation}  
 from the geometry of $\mathcal{H}$ (see Fig.~\ref{Fig:Hex_Proof_2}), it follows from \eqref{Eq:OPE:UB:Sparse:PF} that
\begin{equation}
\begin{aligned}
\!\!\!\!\!\phi(\ell) \leq\! &-\log \Pr\!\Bigg(P_{Y_0}G_{U^*Y_0} \!>\! \frac{\omega}{\theta}\Bigg(\!\!\frac{\sqrt{3}}{2}\sqrt{(D+L)^2 + u}+\\
& L\Bigg)^\alpha\mid \Phi \cap \mathcal{H}\neq \emptyset\Bigg)-
 \log\Pr(\Phi \cap \mathcal{H}\neq \emptyset). 
\end{aligned}\label{Eq:OPE:UB:Sparse:a}
\end{equation}
By inspecting  the scalings of the two terms at the right-hand of  \eqref{Eq:OPE:UB:Sparse:a} as $\ell\rightarrow\infty$, an asymptotic upper bound on $\phi(\ell)$ is obtained as shown in Lemma~\ref{Lem:OPE:UB:Sparse}, which is proved in Appendix~\ref{App:OPE:UB:Sparse}. 
\begin{lemma}\label{Lem:OPE:UB:Sparse}\emph{For sparse scattering and as $\ell\rightarrow\infty$, the OPE for a typical mobile satisfies 
\begin{equation}
\phi(\ell) \preceq \frac{1}{2} \log \ell. 
\end{equation}
}
\end{lemma}

\subsubsection{Main Result and Remarks} The following theorem results  from combining Lemma~\ref{Lem:OPE:LB:Sparse} and Lemma~\ref{Lem:OPE:UB:Sparse}. 

\begin{theorem} \label{Theo:OPE:Sparse}  
\emph{For sparse scattering and as $\ell\rightarrow\infty$,  the OPE $\phi$ for a typical mobile satisfies 
\begin{equation}\label{Eq:PoutExp:Sparse}\eqbox{
\frac{1}{2}\l(1 - \frac{2}{\alpha}\r)\log \ell \preceq \phi(\ell)\preceq   \frac{1}{2}\log \ell. } 
\end{equation} }
\end{theorem}

Several remarks can be made. 

\begin{enumerate}

\item The scaling of the OPE $\phi$ in Theorem~\ref{Theo:OPE:Sparse} is largely determined by the \emph{left-tail probability} [see \eqref{Eq:OPE:LB:Sparse:a} and \eqref{Eq:OPE:UB:Sparse:a}]   of  the  distance $D$ from the typical BS to the boundary of the affiliated cluster. The dominance of $D$ in determining $\phi$ is due to that its distribution has a \emph{linear} left tail [see \eqref{Eq:Dist:D}] that is heavier than the  distribution tails of other random network  parameters.   As can be observed from \eqref{Eq:PoutExp:Sparse}, the asymptotic bounds on $\phi$ are tighter for larger $\alpha$. The reason is that the right tail of the interference-power distribution becomes lighter (with steeper slope)  as $\alpha$ increases, which strengthens the mentioned dominance of $D$ and  thereby tightens  bounds on $\phi$.

\item Theorem~\ref{Theo:OPE:Sparse} shows that  $\phi(\ell)$  scales logarithmically with increasing $\ell$. In contrast, from  Theorem \ref{Theo:CCOPE:Sparse},  the scaling of $\phi^{cc}(\ell)$ for a cluster-center mobile  is much faster, namely  at least sub-linearly with increasing  $\ell$. The reason for this difference in the OPE scaling is that the typical mobile accounts for not only cluster-interior mobiles but also cluster-edge mobiles that are exposed to strong  interference and as a result has much higher  outage probability than a  cluster-interior mobile. This suggests that cluster-edge mobiles are the bottleneck of network coverage and should be protected from strong inter-cluster interference by e.g., applying fractional frequency reuse \cite{Boudreau:InterfCoordCancel4G:2009} along cluster edges.

\item The OPE scaling in Theorem~\ref{Theo:OPE:Sparse} is closely related to the fact that the fraction of mobiles that are near cluster edges is approximately proportional to $\rho^{-1}$ or equivalently $\ell^{-\frac{1}{2}}$. Given the dominance of the outage probabilities for the cluster-edge mobiles over those of the cluster-interior mobiles, the outage probability for the typical mobile is expected to be approximately proportional to the fraction of cluster-edge mobiles and hence $\ell^{-\frac{1}{2}}$. Consequently, the resultant OPE should  be proportional to $\frac{1}{2}\log \ell$,  which matches the result in Theorem~\ref{Theo:OPE:Sparse}.

\item  Unlike  Theorem~\ref{Theo:CCOPE:Sparse} (see Remark $3$), Theorem~\ref{Theo:OPE:Sparse} does not reveal the throughput scaling for a typical mobile. The reason is that  the distribution of the distance  $D$ from a typical mobile to the boundary of the corresponding  cluster [see \eqref{Eq:Dist:D}] dominates the  OPE  but is independent with the outage threshold $\theta$ that determines the throughput. 

\end{enumerate}

\section{OPE with Rich Scattering}\label{Section:OPE:Rich}
Sparse scattering is assumed in the preceding section. In this section, rich scattering is considered and the corresponding OPE is  analyzed for cluster-center and typical mobiles separately. It is shown  that rich scattering decreases the OPE for cluster-center mobiles but has no effect on the OPE for the typical mobiles.

\subsection{OPE for Cluster-Center Mobiles}
\subsubsection{Asymptotic Lower Bound on the OPE}
The presence of rich scattering results in channel fading and hence affects the OPE. In particular, the resultant  distributions of transmission power given channel inversion and interference-channel gains are characterized in Lemma~\ref{Lem:FadTail:Rich} in the sequel.  The effect of rich scattering is reflected in the difference between  Lemma~\ref{Lem:TXPwr:Tail} and Lemma~\ref{Lem:FadTail:Rich}.   

\begin{lemma}\label{Lem:FadTail:Rich} \emph{For rich scattering and an arbitrary BS  $Y\in  \Phi\cap\bar{\mathcal{C}}(T^*, \rho)$, as $x\rightarrow\infty$,  
\begin{equation}
\Pr(P_Y G_{U^* Y} > x) \sim \frac{\omega^N\Gamma\l(\frac{\alpha \nu}{2} + 1\r)\Pr(N = \nu)}{(\pi\lambda)^{\frac{\alpha \nu}{2}}} x^{-\nu}
\end{equation}
where $\Gamma(\cdot)$ denotes the Gamma function. 
}
\end{lemma}

The proof of Lemma~\ref{Lem:FadTail:Rich} is given in Appendix~\ref{App:FadTail:Rich}. Consider the lower bound on the OPE in \eqref{Eq:CCOPE:LB:Sparse} based on the sequence of compound Poisson rvs $\{Z_n\}$, which also holds for $\phi^{cc}$ with rich scattering.  To analyze the scaling of the lower bound as $\ell\rightarrow\infty$, the large deviation   of $Z_n$ is characterized  as follows. 

\begin{lemma} \label{Lem:Zn:Rich}\emph{For rich scattering and as $\ell\rightarrow\infty$, 
\begin{equation}
\Pr\l(Z_n > \ell^\frac{\alpha}{2} x\r) \sim \frac{\omega^N\Gamma\l(\frac{\alpha \nu}{2} + 1\r)\Pr(N = \nu)}{(\pi\lambda)^{\frac{\alpha \nu}{2}}x^{\nu}} \ell^{-\frac{\alpha \nu}{2} +1}   \label{Eq:Zn:Rich}	
\end{equation}
with $n=1, 2, \cdots$. 
}
\end{lemma}
 The proof of Lemma~\ref{Lem:Zn:Rich} can be straightforwardly modified from  that of Lemma~\ref{Lem:LargeDev:Z} by applying   Lemma~\ref{Lem:FadTail:Rich} in place of  Lemma~\ref{Lem:TXPwr:Tail}; the details are omitted for brevity.  It can be observed from  \eqref{Eq:Zn:Rich} that  the distribution of $Z_n$ does not have a sub-exponential tail as for the case with sparse scattering. This makes it difficult to apply the contraction principle  as before to derive  the scaling of the series $\sum_{n=1}^\infty n^{-\frac{\alpha}{2}} Z_n$,  
 which is needed for obtaining an asymptotic lower bound on $\phi$. To overcome this difficulty, the current analysis applies  the following result from  \cite[Theorem~$2.3$]{Cline:SeriesRandomVariableRegVaryTails:1983}. 

\begin{lemma}[\cite{Cline:SeriesRandomVariableRegVaryTails:1983}]\label{Lem:Sum:RV} \emph{
Consider a sequence of i.i.d. rvs $\{\tilde{Z}_n\}$ whose distribution belongs to  $\mathsf{RV}(\tau)$ with $\tau > 0$ and 
a sequence of nonnegative scalars $\{\rho_n\}$  with  $\sum_{n=1}^\infty \rho_n^v$ being  finite  for some $0 < v < \min(1, \tau)$. The tail probability of $\sum_{n=1}^\infty \rho_n \tilde{Z}_n$ scales as 
\begin{equation}
\Pr\l(\sum_{n=1}^\infty \rho_n \tilde{Z}_n  > x\r) \sim \sum_{n=1}^\infty \rho_n^\tau\Pr(\tilde{Z}_n > x), \ \  x \rightarrow \infty.
\end{equation}}
\end{lemma}

Based on Lemma~\ref{Lem:Zn:Rich} and Lemma~\ref{Lem:Sum:RV}, it is proved in Appendix~\ref{App:CCOPE:LB:Rich} that as $\ell\rightarrow\infty$, the OPE can be upper bounded as shown in the following lemma. 
\begin{lemma}\label{Lem:CCOPE:LB:Rich}\emph{For rich scattering and as $\ell\rightarrow\infty$, the OPE for a  cluster-center mobile satisfies 
\begin{equation}
\phi^{cc}(\ell) \succeq \l(\frac{1}{2}\alpha \nu -1\r)\log \ell. \label{Eq:CCOPE:LB:Rich}
\end{equation}
}
\end{lemma}

\subsubsection{Asymptotic Upper Bound on the OPE} 
The following lemma is proved using Lemma~\ref{Lem:FadTail:Rich} and applying  a procedure similar to that for   proving Lemma~\ref{Lem:CCOPE:UB:Sparse} with  the details omitted to keep the exposition precise. 

\begin{lemma}\label{Lem:CCOPE:UB:Rich}\emph{For rich scattering and as $\ell\rightarrow\infty$, the OPE for a cluster-center mobile satisfies 
\begin{equation}
\phi^{cc}(\ell) \preceq \frac{1 }{2}\alpha \nu \log \ell. \label{Eq:CCOPE:UB:Rich}
\end{equation}}
\end{lemma}

\subsubsection{Main Result and Remarks} The following theorem follows directly from Lemma~\ref{Lem:CCOPE:LB:Rich} and \ref{Lem:CCOPE:UB:Rich}. 

\begin{theorem} \label{Theo:CCOPE:Rich} 
\emph{For    rich scattering and  as  $\ell\rightarrow\infty$, the OPE for  a cluster-center  mobile satisfies 
\begin{equation}\eqbox{
\l(\frac{1}{2} \alpha \nu-1\r)\log \ell\preceq \phi^{cc}(\ell) \preceq \frac{1}{2} \alpha \nu\log \ell
}
\end{equation}
where $\nu$ is the minimum signal diversity order. 
}
\end{theorem}

A few remarks are in order. 
\begin{enumerate}
\item By comparing Theorem \ref{Theo:CCOPE:Rich} with Theorem~\ref{Theo:CCOPE:Sparse}, one can see that channel fading caused by rich scattering degrades $\phi^{cc}$ dramatically. To be specific, as $\ell\rightarrow\infty$, $\phi^{cc}(\ell)$ can scale at least   \emph{sub-linearly} with $\ell$ for sparse scattering but only \emph{logarithmically} for rich scattering.  Roughly speaking, fading  increases the randomness in interference   and thereby reduces the level   of spatial separation. This introduces a larger  number of significant interferers for the cluster-center mobiles with respect to the case of no fading and hence compromises the effectiveness  of MCC. This is the key reason for the slower OPE scaling  in   Theorem \ref{Theo:CCOPE:Rich} compared with that in Theorem~\ref{Theo:CCOPE:Sparse}. 

\item For single-cell transmissions over fading channels, increasing the BS density does not change the  outage probability for an interference-limited network, as shown in \cite{Andrews:TractableApproachCoverageCellular:2010}. In contrast, Theorem \ref{Theo:CCOPE:Rich} indicates that  it is possible to reduce outage probability by deploying more BS so long as the numbers of cooperating  BSs  increase proportionally. 

\item It is well-known that the effect  of fading can be alleviated by diversity techniques \cite{TseBook:WirelessComm:05}. This is reflected in Theorem~\ref{Theo:CCOPE:Rich} where $\phi^{cc}$ is observed to increase approximately  linearly with the minimum diversity order $\nu$  if $\alpha \nu$ is large. For this case, the asymptotic bounds on $\phi^{cc}$ are observed to be tight. Moreover, $\phi^{cc}$ also grows approximately proportionally with increasing  $\alpha$ as inter-cluster interference is more severely  attenuated. 

\end{enumerate}

\subsection{OPE for Typical  Mobiles} The type of scattering has no effect on the scaling of OPE for a typical mobile as $\ell\rightarrow\infty$ as stated in the following theorem.   
\begin{theorem} \label{Theo:OPE:Rich}  
\emph{For rich scattering and as $\ell\rightarrow\infty$,  the OPE  for the typical mobile scales  as shown in Theorem~\ref{Theo:OPE:Sparse}. }
\end{theorem}

The proof of Theorem~\ref{Theo:OPE:Rich} can be easily modified from that of Theorem~\ref{Theo:OPE:Sparse} based on the new distribution of the rvs $\{P_YG_{U^*Y}\}$ in Lemma~\ref{Lem:FadTail:Rich}.  The detailed proof of Theorem~\ref{Theo:OPE:Rich} is omitted. 

The insensitivity  of $\phi$  with respect to the change on the scattering environment is due to that the distribution of  $D$ is independent with scattering and has a dominant effect on $\phi$ compared with the distributions of other network parameters (see Remark~1 on Theorem~\ref{Theo:OPE:Sparse}). Furthermore, since the distribution  function of $D$ is also independent with the diversity order $N$, it can be observed   by comparing  Theorem~\ref{Theo:CCOPE:Rich} and \ref{Theo:OPE:Rich} that unlike a cluster-center mobile, a typical mobile   does not benefit from transmit diversity  for improving the OPE scaling. Therefore, the result in Theorem~\ref{Theo:OPE:Rich} reiterates the importance of suppressing  inter-cluster interference for cluster-edge mobiles  to improve network coverage via MCC.

\begin{figure*}
\begin{center}
\subfigure[Sparse scattering ]{\includegraphics[width=12cm]{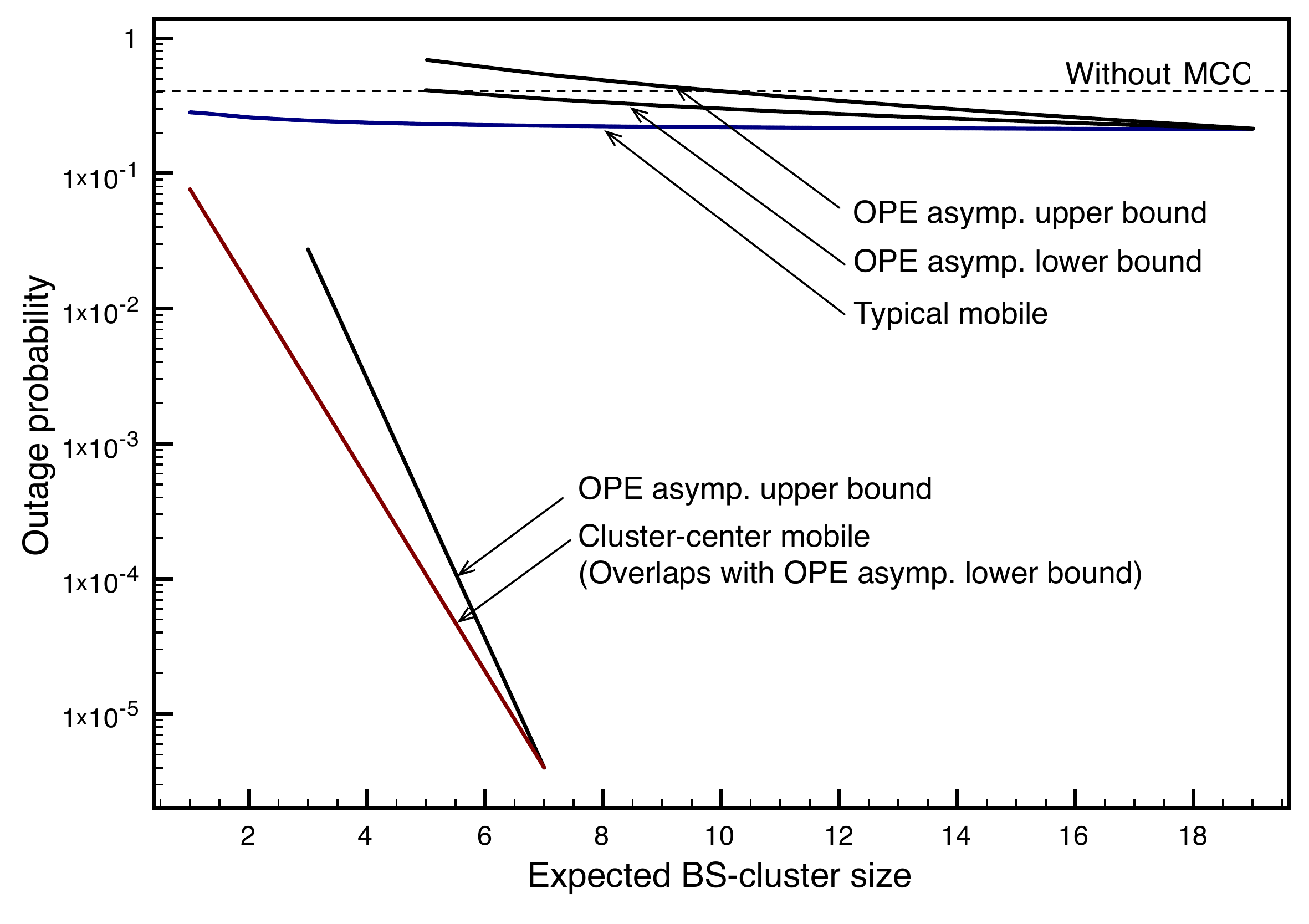}}\\
\subfigure[Rich scattering ]{\includegraphics[width=12cm]{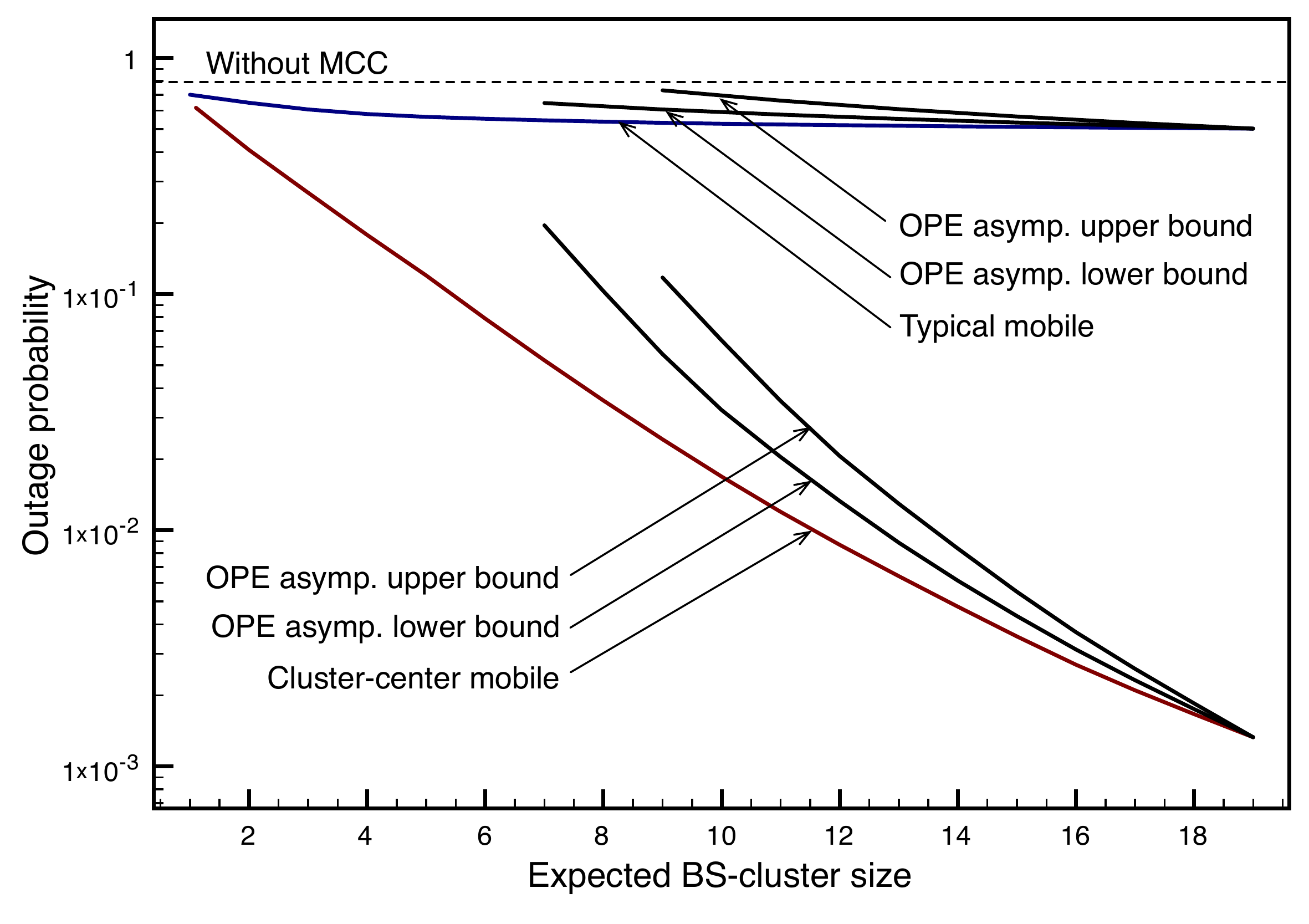}}
\caption{Outage probability versus expected BS-cluster size $\ell$ with (a) sparse scattering or  (b) rich scattering. For comparison, the outage probability for the case of no MCC is specified by dashed horizontal lines. }
\label{Fig:Pout}
\end{center}
\end{figure*}

\section{Simulation Results}\label{Section:Simulations}
The simulation method and settings are summarized as follows. The infinite network region is approximated by a disk centered at the origin, where  BSs are  Poisson distributed  with density $\lambda=10^{-2}$ and the disk area is chosen such that the expected number of BSs in the disk is  $200$ i.e., the disk area is $200/\lambda=2\times 10^4$. The typical cluster region is centered at the origin and the size is determined by the expected BS-cluster size  $\ell$. { The main and side lobes of beams are uniformly distributed in the intervals $[\delta, \delta'] = [6, 10]$ and $[0, \gamma] = [0, 1]$, respectively.} Other  parameters are sets as  $\alpha = 4$,  $N=3$ (for rich scattering), and   $\theta = 3$.

In Fig.~\ref{Fig:Pout}, outage probability is plotted against increasing $\ell$ for different combinations of  sparse/rich scattering and a cluster-center/typical mobile. To evaluate the asymptotic results derived in the preceding sections,   Fig.~\ref{Fig:Pout} also displays curves obtained from the asymptotic bounds on the OPE as follows. Consider a typical mobile and let $\phi^{+}$ and $\phi^{-}$ represent the asymptotic upper and lower bounds on the OPE, respectively. Note that  outage probability can be approximated as  $\Pout  \approx b e^{-\phi(\ell)}$ if $\ell \gg 1$ where $b$ is a constant. For this reason,  the functions  $b_1e^{-\phi^{+}(\ell)}$ and $b_2e^{-\phi^{-}(\ell)}$ are plotted in Fig.~\ref{Fig:Pout} and identified by the legends ``OPE asymptotic upper bound" and ``OPE asymptotic lower bound", respectively, where the constants $b_1$ and $b_2$ are chosen such that  the matching analytical and  simulation curves overlap at their rightmost points for ease of comparison. Similar curves are also plotted in  Fig.~\ref{Fig:Pout} for a cluster-center mobile. The curves based on analysis and simulation are observed to be closely aligned if $\ell$ is sufficiently large,  indicating that the derived  asymptotic bounds on the OPE (especially the asymptotic lower bound) are  accurate. In particular, for the cluster-center mobile with sparse scattering, the curve from the asymptotic lower bound on the OPE overlaps with the simulation curve and hence this bound is tight even  for small values of $\ell$.
  
Next, it can be observed from Fig.~\ref{Fig:Pout} that as $\ell$ increases, the outage probability for a cluster-center  mobile  decreases rapidly but the outage probability for a typical mobile remains almost unchanged and close to the result for the case of no MCC (specified in Fig.~\ref{Fig:Pout} using dashed lines). In other words, it is verified that MCC benefits only cluster-interior mobiles and cluster-edge mobiles limit network coverage. This observation is consistent with findings from implementing MCC in practical networks {\cite{Barb:CoMPHeterNet:2012, Annapureddy:CoordinatedTXWWAN:2010, Irmer:CoMPConceptsPerformFieldTrial:2011}}. Furthermore, with respect to sparse scattering, rich scattering is observed to increase outage probability for cluster-center mobiles by up to several orders of magnitude. 

\begin{figure*}
\begin{center}
\includegraphics[width=12cm]{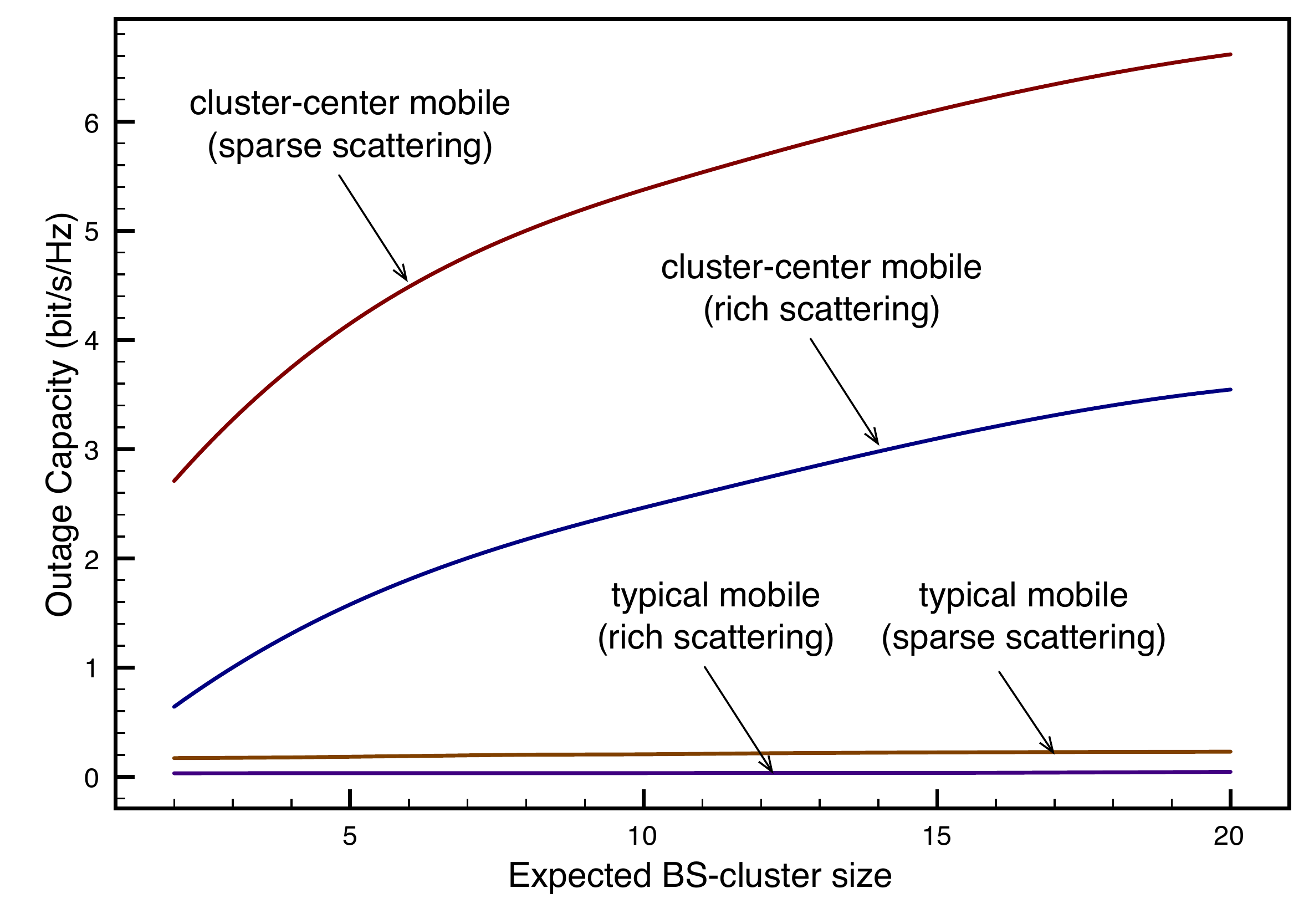}
\caption{Outage capacity per mobile  versus expected BS-cluster size $\ell$ for the maximum outage probability of $0.05$. Note that the outage capacity per mobile for the case of no MCC is approximately zero.  }
\label{Fig:Capacity}
\end{center}
\end{figure*}

Fig.~\ref{Fig:Capacity} compares the outage capacity of cluster-center and typical  mobiles, namely their maximum throughput given the maximum  outage probability of   $0.05$. The observations from Fig.~\ref{Fig:Capacity} agree   with those from Fig.~\ref{Fig:Pout}. Specifically, the outage capacity for a typical mobile is marginal even as $\ell$ increases while the  capacity for mobiles without MCC is approximately zero. In contrast, the outage capacity for a cluster-center mobile increases rapidly with growing $\ell$ and sparse scattering results in much higher capacity than rich scattering.  

\section{Concluding Remarks}\label{Section:Conclusion}
In this paper, a novel  model was  proposed for a  cellular downlink  network with MCC. The network coverage was analyzed in terms of  the outage-probability exponent. It was shown that though the performance gain for  cluster-interior mobiles from MCC is large, the gain for a typical mobile is small  as it is likely to be located near the edge of a base-station cluster and exposed to strong inter-cluster  interference. This finding  provides an explanation for the marginal gain  of MCC in practice, and suggests the need to design a new medium-access-control protocol or apply fractional-frequency reuse for  protecting cluster-edge mobiles. 

This work opens  several interesting directions for further research. In particular,   instead of using a lattice, base-stations can be clustered by a random process such as a Poisson random tessellation that gives non-uniform expected BS-cluster sizes.  Moreover, the current  interference-coordination algorithm that requires multi-antennas at base stations can be replaced with a network-MIMO algorithm that supports  cooperation  between single-antenna base stations at the cost of inter-cell data exchange. Last, the proposed analytical framework  can be applied to study the performance of other MCC algorithms and    heterogeneous networks with MCC. 

\appendices  

\section{Proof of  Lemma~\ref{Lem:TXPwr:Tail}}\label{App:TXPwr:Tail} Consider  an arbitrary  BS $Y\in \Phi$. For convenience, define $\beta = W_Y/G_{U^* Y}$ with support $[\delta/\gamma, \infty)$ and the probability density function is denoted as $f_\beta$. Using  $P_Y = \omega L_Y^\alpha/W_Y$ from \eqref{Eq:SigPwr:Sparse}, 
\begin{equation}\nn
\Pr( P_Y G_{U^* Y} > x) = \int_{\frac{\delta}{\gamma}}\Pr\l(L_Y^\alpha > \frac{\tau x }{\omega}\r)f_{\beta}(\tau) d\tau,\qquad x > 0. 
\end{equation}
Substituting the distribution function of $L_Y$ in \eqref{Eq:L:Dist} gives 
\begin{equation}
\Pr( P_Y G_{U^* Y} > x) =\int^\infty_{\frac{\delta}{\gamma}} e^{-\pi\lambda \l(\tau x/\omega\r)^{\frac{2}{\alpha}}}f_\beta(\tau)d\tau.  \label{Eq:Proof:c1a}
\end{equation}
The right-hand size of \eqref{Eq:Proof:c1a} can be expanded for $\epsilon > 0$ as 
\begin{equation}
\begin{aligned}
\!\!\!\!\Pr( P_Y G_{U^* Y} > x)
&= \int^\infty_{(1+\epsilon)\frac{\delta}{\gamma}} e^{-\pi\lambda \l(\tau x/\omega\r)^{\frac{2}{\alpha}}}f_\beta(\tau)d\tau +  \\ 
&\qquad  \int^{(1+\epsilon)\frac{\delta}{\gamma}}_{\frac{\delta}{\gamma}} e^{-\pi\lambda \l(\tau x/\omega\r)^{\frac{2}{\alpha}}}f_\beta(\tau)d\tau. 
\end{aligned}
\label{Eq:Proof:c1}
\end{equation}
It follows that 
\begin{align}
\Pr( P_Y G_{U^* Y} > x) 
&\geq e^{-\pi\lambda \l((1+\epsilon)\frac{\delta x}{\gamma\omega} \r)^{\frac{2}{\alpha}}}\times\nn\\
&\qquad \qquad \Pr\l(\frac{\delta}{\gamma}\leq \beta <  {(1+\epsilon)\frac{\delta}{\gamma}}\r). \nn
\end{align}
Thus, as $x \rightarrow\infty$, 
\begin{equation}
-\log \Pr(P_Y G_{U^* Y} > x) \preceq (1+\epsilon)^{\frac{2}{\alpha}}\pi\lambda \l(\frac{\delta }{\gamma\omega} \r)^{\frac{2}{\alpha}}x ^{\frac{2}{\alpha}}. \label{Eq:Proof:c3}
\end{equation}
Next, it can be obtained form \eqref{Eq:Proof:c1} that 
\begin{align}
\!\!\!\!\Pr( P_Y G_{U^* Y} > x) 
&\leq e^{-\pi\lambda \l((1+\epsilon)\frac{\delta x}{\gamma\omega} \r)^{\frac{2}{\alpha}}}\Pr\l(\beta > {(1+\epsilon)\frac{\delta}{\gamma}}\r) + \nn\\
&\quad   e^{-\pi\lambda \l(\frac{\delta x}{\gamma\omega} \r)^{\frac{2}{\alpha}}}\Pr\l(\frac{\delta}{\gamma}\leq \beta \leq 
(1+\epsilon)\frac{\delta}{\gamma}\r). \nn
\end{align}
As a result, 
\begin{equation}
-\log \Pr( P_Y G_{U^* Y} > x) \succeq \pi\lambda \l(\frac{\delta }{\gamma\omega} \r)^{\frac{2}{\alpha}}x ^{\frac{2}{\alpha}}, \ \ x \rightarrow\infty. \label{Eq:Proof:c2}
\end{equation}
Combining  \eqref{Eq:Proof:c2} and \eqref{Eq:Proof:c3} and letting  $\epsilon \rightarrow 0$ yield  the desired result in \eqref{Eq:PG:Tail}. Last, the claim  of $\E[P_Y G_{U^*Y}]$ being finite follows from the fact that $\E[L^\alpha]$, $\E[G_{U^*Y}]$ and $\E[W^{-1}_Y]$ are all bounded, which follows from the distribution of $L$ in \eqref{Eq:L:Dist} and those of $G_{U^*Y}$ and $W_Y$ in Lemma~\ref{Lem:Jindal}. \hfill $\blacksquare$


\section{Proof of Lemma~\ref{Lem:LargeDev:Z}}
\label{App:LargeDev:Z} First, consider the case of $\alpha > 4$. Since  $P_Y G_{U^* Y}\in \mathsf{WE}(\tau)$ with $0 < \tau < 0.5$ according to Lemma~\ref{Lem:TXPwr:Tail} and $\E[Z_n] = \ell\E[P_YG_{U^* Y}]$,  applying Lemma~\ref{Lem:CompPoisson}  gives  the desired result in \eqref{Eq:LogZn:LB:a}. 

Next, consider the case of $2 < \alpha \leq 4$. It is claimed that  as $\ell \rightarrow\infty$, 
\begin{equation}
\begin{aligned}
-\log&\Pr(Z_n > \ell^{\frac{\alpha}{2}} x ) \succeq \\
&-\log\Pr\l(\sum_{Y\in \Phi\cap\mathcal{A}_n} \l(P_YG_{U^*Y}\r)^{\frac{4(1+\epsilon)}{\alpha}} > \ell^{\frac{\alpha}{2}} x \r)
\end{aligned}
\label{Eq:Lem:Zn:4}
\end{equation}
with $\epsilon, x > 0$. To prove this claim,  let $V_1, V_2, \cdots, V_k$ denote $k$ i.i.d. rvs following the same distribution as $P_YG_{U^*Y}$ for an arbitrary $Y\in \Phi\cap\bar{\mathcal{C}}(T^*, \rho)$.  By using Lemma~\ref{Lem:TXPwr:Tail} and applying the \emph{contraction principle} from large-deviation theory \cite[Theorem~$4.2.1$]{DemboBoo:LargeDeviation}, for a set of nonnegative numbers $\{x_1, x_2, \cdots, x_\ell\}$ and as $x\rightarrow\infty$, 
\begin{align}
-\log\Pr\l(\sum_{n=1}^k V_k > x\r) &\sim \pi\lambda\l(\frac{ \delta}{\gamma\omega}\r)^{\frac{2}{\alpha}} \inf_{\sum_{n=1}^k x_n > x}\sum_{n=1}^k x_k^{\frac{2}{\alpha}} \nn\\
&  \sim \pi\lambda\l(\frac{ \delta}{\gamma\omega}\r)^{\frac{2}{\alpha}}x^{\frac{2}{\alpha}} \label{Eq:Lem:Zn:1}
\end{align}
where \eqref{Eq:Lem:Zn:1} results from the inequality $\l(\sum_{n=1}^kx_k\r)^p \leq   \sum_{n=1}^kx_k^p$ if $ 0 \leq p \leq 1$. It follows from Lemma~\ref{Lem:TXPwr:Tail} that as $x \rightarrow\infty$, 
\begin{equation}
-\log\Pr\l(V_1^{\frac{4(1+\epsilon)}{\alpha}} > x\r) \sim  \pi\lambda\l(\frac{ \delta}{\gamma\omega}\r)^{\frac{2}{\alpha}}x^{\frac{1}{2(1+\epsilon)}}. \label{Eq:Lem:Zn:2a}
\end{equation}
Using \eqref{Eq:Lem:Zn:2a} and again applying the contraction principle give
\begin{equation}
-\log\Pr\l(\sum_{n=1}^k V_k^{\frac{4(1+\epsilon)}{\alpha}} > x\r) \sim  \pi\lambda\l(\frac{ \delta}{\gamma\omega}\r)^{\frac{2}{\alpha}}x^{\frac{1}{2(1+\epsilon)}}\label{Eq:Lem:Zn:2} 
\end{equation}
as $x \rightarrow\infty$. 
Given $2 < \alpha \leq 4$,  comparing  \eqref{Eq:Lem:Zn:1} and \eqref{Eq:Lem:Zn:2} yields 
\begin{equation}
\!-\log\Pr\l(\sum_{n=1}^k V_k > x\r) \!\succeq\! -\log\Pr\l(\sum_{n=1}^k V_k^{\frac{4(1+\epsilon)}{\alpha}} \!>\! x\r) \label{Eq:Lem:Zn:3} 
\end{equation}
as $x \rightarrow\infty$. 
Since the inequality in \eqref{Eq:Lem:Zn:3} holds for arbitrary $k$,  the claimed inequality in \eqref{Eq:Lem:Zn:4} is proved. Recall that  $P_Y G_{U^* Y}$ for arbitrary  
$Y\in  \Phi\cap\bar{\mathcal{C}}(T^*, \rho)$ has the same distribution as $V_1$. By inspecting  \eqref{Eq:Lem:Zn:2a},  $P_Y G_{U^* Y}\in \mathsf{WE}(\tau)$ with $0< \tau < 0.5$. Therefore, it can be derived similarly as \eqref{Eq:LogZn:LB:a} in the lemma statement that 
\begin{equation}\label{Eq:Lem:Zn:5}
\begin{aligned}
-\log \Pr&\l(\sum_{Y\in \Phi\cap \mathcal{A}_n} (P_Y G_{U^*Y})^{\frac{4(1+\epsilon)}{\alpha}} > \ell^{\frac{\alpha}{2}} x\r) \\
&\sim  \pi\lambda\l(\frac{ \delta}{\gamma\omega}\r)^{\frac{2}{\alpha}}\ell^{\frac{\alpha}{4(1+\epsilon)}}x^{\frac{1}{2(1+\epsilon)}}. 
\end{aligned}
\end{equation}
Substituting \eqref{Eq:Lem:Zn:5} into \eqref{Eq:Lem:Zn:4} and letting $\epsilon \rightarrow 0$ gives \eqref{Eq:LogZn:LB:b} in the lemma statement. \hfill $\blacksquare$

\section{Proof of Lemma~\ref{Lem:CCOPE:UB:Sparse}}
\label{App:CCOPE:UB:Sparse}

From the definition of $Z_{\epsilon}$ in \eqref{Eq:Z:Eps} and \eqref{Eq:CCOPE:UB:Sparse:a}, 
\begin{equation}
\begin{aligned}
\!\!\!\!\phi^{cc}(\ell)  
\leq -\log\Bigg(\!\!\Pr&\Bigg(\!\!\Bigg(\!\frac{2}{\sqrt{3}}(1+\epsilon)\rho\Bigg)^{-\alpha}P_YG_{T^*  Y} \!>\! \frac{\omega}{\theta} \mid \\
& Y\in \Phi\cap \mathcal{A}_\epsilon\!\Bigg)\Pr(\Phi\cap \mathcal{A}_\epsilon\neq \emptyset)\!\Bigg).  \label{Eq:Pout:LB:LargeAlp}
\end{aligned}
\end{equation}
Since $\Pr(\Phi\cap \mathcal{A}_\epsilon\neq \emptyset) = \l(1 - e^{-\epsilon \ell}\r)$ and $\ell = 2\sqrt{3}\rho^2\lambda$, using \eqref{Eq:Pout:LB:LargeAlp} and  applying Lemma~\ref{Lem:TXPwr:Tail}  give that 
\begin{equation}
\phi^{cc}(\ell) \preceq \frac{2\pi(1+\epsilon)^2}{3\sqrt{3}}\l(\frac{ \delta}{\theta\gamma}\r)^{\frac{2}{\alpha}} \ell, \qquad \ell \rightarrow\infty. \label{Eq:OPExp:LB}
\end{equation}
The desired result follows  from   \eqref{Eq:OPExp:LB} by  letting $\epsilon \rightarrow 0$. \hfill $\blacksquare$

\section{Proof of Lemma~\ref{Lem:PoutUB:Sparse:a}}
\label{App:PoutUB:Sparse:a}
By expanding outage probability defined in \eqref{Eq:Pout:Def}, 
\begin{equation}
\Pout \leq \Pr\l( I >  \frac{\omega}{\theta} \mid L \leq \frac{D}{2}\r) + \Pr\l(L >  \frac{D}{2}\r). 
\end{equation}
The substitution of  \eqref{PF:T2:Eq2} yields that
\begin{align}
\Pout &\leq \Pr\Bigg(\sum_{n=1}^\infty Z_n  \l(\max(\sqrt{n}\rho - \rho + D-L, L)\r)^{-\alpha} > \frac{\omega}{\theta}\mid \nn\\
&\qquad\qquad  L \leq \frac{D}{2}\Bigg) + \Pr\l(L >  \frac{D}{2}\r)\nn\\
&
\begin{aligned}
&\leq\Pr\l(\sum_{n=1}^\infty Z_n  \l(\sqrt{n}\rho - \rho + \frac{D}{2}\r)^{-\alpha} > \frac{\omega}{\theta} \r) +\\
&\qquad \qquad  \Pr\l(L >  \frac{D}{2}\r). 
\end{aligned}\label{PF:T2:Eq3}
\end{align}
Since $D \leq \rho$, the replacement of $\rho$ in \eqref{PF:T2:Eq3} with $D/2$ further upper bounds $\Pout$ as 
\begin{equation}
\Pout\leq \Pr\l(\sum_{n=1}^\infty n^{-\frac{\alpha}{2}} Z_n > \frac{D^{\alpha}\omega}{2^\alpha \theta}\r)+\Pr\l(L >  \frac{D}{2}\r). \label{PF:T2:Eq4:a}
\end{equation}
Applying the similar method as for obtaining \eqref{PF:T2:Eq3} results in an upper bound on the first term on the right-hand side of \eqref{PF:T2:Eq4:a}:
\begin{equation}\nn
\!\!\!\!\!\!\!\begin{aligned}
&\Pr\l(\sum_{n=1}^\infty n^{-\frac{\alpha}{2}} Z_n > \frac{D^{\alpha}\omega}{2^\alpha \theta}\r)\leq \min_{z > 0} \Bigg[\Pr\Bigg(\sum_{n=1}^\infty n^{-\frac{\alpha}{2}} Z_n > \\
& \qquad \qquad \frac{\omega z}{2^{\alpha}\theta}\Bigg) + \Pr(D^{\alpha} \leq z)\Bigg].  
\end{aligned}
\end{equation}
By  combining \eqref{PF:T2:Eq4:a} and the last inequality, 
\begin{equation}\label{Eq:Pout:UB:PF}
\begin{aligned}
\Pout\leq \min_{z > 0} \Bigg[\Pr&\l(\sum_{n=1}^\infty n^{-\frac{\alpha}{2}} Z_n >  \frac{\omega z}{2^{\alpha}\theta}\r) \\
&+ \Pr(D^{\alpha} \leq z)\Bigg]+\Pr\l(L >  \frac{D}{2}\r). 
\end{aligned}
\end{equation}
 As $\ell\rightarrow\infty$, the term on the right-hand size of  \eqref{Eq:Pout:UB:PF} that decays at the slowest  rate dominates the other two terms. Specifically, given  \eqref{Eq:Pout:UB:PF} and the definition of $\phi$ in \eqref{Eq:OPE:Def:a}, 
applying \cite[Lemma~$1.2.15$]{DemboBoo:LargeDeviation}  yields the desired result in the lemma statement. \hfill $\blacksquare$

\section{Proof of Lemma~\ref{Lem:OPE:LB:Sparse}}
\label{App:OPE:LB:Sparse} Consider the three terms in the  asymptotic lower bound on $\phi$ in  Lemma~\ref{Lem:PoutUB:Sparse:a}. 
By setting  $2^{-\alpha}\theta^{-1} \omega z = \E[Z_n] +  \rho^{2(1+\epsilon)}$ with $\epsilon > 0$, the procedure similar to that  for obtaining  Lemma~\ref{Lem:LargeDev:Z} can be applied to derive the following asymptotic lower bound on the first term: as $\ell \rightarrow\infty$, 
\begin{equation}
\begin{aligned}
-\log \Pr\Bigg(\sum_{n=1}^\infty n^{-\frac{\alpha}{2}} &Z_n >  \frac{\omega z}{2^{\alpha}\theta }\Bigg) \succeq \\
& \pi\lambda\l(\frac{ \delta}{\gamma\omega}\r)^{\frac{2}{\alpha}} \l(\frac{\ell}{2\sqrt{3}\lambda}\r)^{\frac{1+\epsilon}{2}}.  
   \end{aligned}
\label{PF:S1:Asymp}
\end{equation}
The scaling of the second term is obtained using \eqref{Eq:Dist:D} and the aforementioned constraint on $y$ as
\begin{equation}\label{PF:S2:Asymp}
-\log \Pr(D^{\alpha} \leq z) \sim \frac{1}{2}\l(1-\frac{2(1+\epsilon)}{\alpha}\r)\log \ell
\end{equation}
as $\ell\rightarrow\infty$. 
Using the distributions of $L$ and $D$ in \eqref{Eq:L:Dist} and \eqref{Eq:Dist:D} respectively, 
\begin{align}
\Pr\l(L >  \frac{D}{2}\r) &=\frac{2}{\rho} \int_0^\rho e^{-\frac{\pi\lambda \tau^2}{4}} \l(1 - \frac{\tau}{\rho}\r)d\tau\nn\\
&\sim \frac{2}{\sqrt{\lambda}\rho}  - \frac{4}{\pi\lambda\rho^2}, \qquad \rho \rightarrow\infty.  \nn
\end{align}
By substituting $\ell = 2\sqrt{3}\rho^2$, the third term scales as 
\begin{equation}
-\log \Pr\l(L >  \frac{D}{2}\r)\sim \frac{1}{2}\log \ell, \qquad \ell \rightarrow\infty. \label{Eq:S3:Asymp} 
\end{equation}
Last, the substitution of \eqref{PF:S1:Asymp}, \eqref{PF:S2:Asymp} and \eqref{Eq:S3:Asymp} into the asymptotic lower bound on $\phi$ in Lemma~\ref{Lem:PoutUB:Sparse:a}  and letting $\epsilon \rightarrow 0 $ lead to   the result in the lemma statement. \hfill $\blacksquare$

\section{Proof of Lemma~\ref{Lem:OPE:UB:Sparse}}
\label{App:OPE:UB:Sparse}
As the area of $\mathcal{H}$ is $u/\sqrt{3}$, the number of BSs in $\mathcal{H}$, namly $|\Phi\cap\mathcal{H}|$,  follows the Poisson distribution with mean $ u\lambda/\sqrt{3}$.  Using this fact and \eqref{Eq:OPE:UB:Sparse:a}, 
\begin{align}
\phi(\ell)  
&\leq  -\log \Bigg(\Pr\Bigg(P_{Y_0}G_{U^*Y_0} > \frac{\omega}{\theta}\Bigg(\frac{\sqrt{3}}{2}\sqrt{(D+L)^2 + u} +\nn\\
&\qquad\qquad   L\Bigg)^\alpha\!\mid\! D < x\!\Bigg)\Pr(D < x)\!\Bigg) -\log(1\!-\!e^{-\frac{u\lambda}{\sqrt{3}}})  \nn\\
&\leq  -\log \Pr\Bigg(P_{Y_0}G_{U^*Y_0} > \frac{\omega}{\theta}\Bigg(\frac{\sqrt{3}}{2}\sqrt{(x+L)^2 + u} +\nn\\
&\qquad\qquad   L\Bigg)^\alpha\Bigg)- \log\Pr(D < x)  -\log(1\!-\!e^{-\frac{u\lambda}{\sqrt{3}}})  \nn
\end{align}
with $x > 0$.  By keeping $u$ and   $x $  constant and letting  $\rho\rightarrow \infty$,  
\begin{align}
\phi(\ell) &\preceq -\log \Pr(D < x) \nn\\
&\sim \log \rho \label{PF:T2:Eq8}
\end{align}
where \eqref{PF:T2:Eq8} uses  the distribution  of $D$ in \eqref{Eq:Dist:D}.  
The substitution of $\ell=2\sqrt{3}\lambda \rho^2$ into \eqref{PF:T2:Eq8} proves the desired result. \hfill $\blacksquare$

\section{Proof of Lemma~\ref{Lem:FadTail:Rich}}
\label{App:FadTail:Rich}
Consider an arbitrary BS $Y \in \Phi$ and the corresponding parameters $\{P_Y, G_Y, W_Y, L_Y\}$. The subscripts of these parameters are omitted in the remainder of the proof to simplify notation. Given $P = \omega L^\alpha/W$ from channel inversion and $x > 0$, it follows from  Lemma~\ref{Lem:Jindal} that 
\begin{align}
\!\!\!\!\!\Pr(PG > x) &= \Pr\l(G  >  \omega^{-1}W L^{-\alpha}x\r)\nn\\
&= \E\l[e^{- \omega^{-1}W L^{-\alpha}x}\r]  \label{Eq:Proof:d1c}\\
&= \E\l[\int_0^\infty e^{- \omega^{-1}x L^{-\alpha} \tau} \frac{\tau^{N-1}}{(N-1)!} e^{-\tau}d\tau\r] \label{Eq:Proof:d1a}\\
&= \E\l[ \frac{1}{(1+  \omega^{-1}x L^{-\alpha})^{N}}\r]\label{Eq:Proof:d1}\\
&\geq \E\l[ \frac{1}{(1+  \omega^{-1}x L^{-\alpha})^{N}}\mid  \frac{x}{ \omega L^{\alpha}} > \log x\r]\times \nn\\
&\qquad\qquad  \Pr\l(\frac{x}{ \omega L^{\alpha}} > \log x\r)\nn\\
&=\E\l[ \frac{\omega^NL^{\alpha N}}{(1+ \frac{\omega L^{\alpha}}{x })^{N}x^N}\mid \frac{x}{ \omega L^{\alpha}} > \log x\r]\times \nn\\
&\qquad\qquad  \Pr\l(\frac{x}{ \omega L^{\alpha}} > \log x\r)\nn\\
&\geq \E\l[ \frac{\omega^NL^{\alpha N}}{\l(1+ \frac{1}{\log x}\r)^{N} x^N} \mid \frac{x}{ \omega L^{\alpha}} > \log x\r]\times \nn\\
&\qquad \qquad \Pr\l(\frac{x}{ \omega L^{\alpha}} > \log x\r)\nn\\
&= \E\l[\frac{\int_0^{\l(\frac{x}{\omega\log x}\r)^{\frac{1}{\alpha}}}\!\!\!\! y^{\alpha N} e^{-\pi \lambda y^2} d(\pi\lambda y^2)}{\l(1+ \frac{1}{\log x}\r)^{N} x^{N}}\r] \label{Eq:Proof:d1b}\\
&\sim \E\l[\frac{\Gamma\l(\frac{\alpha N}{2}+1\r)}{(\pi\lambda)^{\frac{\alpha N }{2}}}x^{- N }\r]\omega^N, \qquad x \rightarrow\infty\nn\\
&\sim \frac{\omega^N\Gamma\l(\frac{\alpha \nu}{2}+1\r)\Pr(N=\nu)}{(\pi\lambda)^{\frac{\alpha \nu }{2}}}x^{- \nu } \label{Eq:Proof:d3}
\end{align}
where \eqref{Eq:Proof:d1c},   \eqref{Eq:Proof:d1a} and \eqref{Eq:Proof:d1b}  use the distributions of $G$ and $W$ in Lemma~\ref{Lem:Jindal}  and $L$ in \eqref{Eq:L:Dist}, respectively. 
Next, from \eqref{Eq:Proof:d1}, 
\begin{align}
\!\!\Pr(PG > x) &\leq \E\l[ \omega^Nx^{-N} L^{\alpha N}\r]\nn \\
&\sim  \frac{\omega^N\Gamma\l(\frac{\alpha \nu}{2}+1\r)\Pr(N=\nu)}{(\pi\lambda)^{\frac{\alpha \nu }{2}}}x^{- \nu } \label{Eq:Proof:d2}
\end{align}
as $x \rightarrow\infty$, where \eqref{Eq:Proof:d2}  is obtained using a similar procedure as \eqref{Eq:Proof:d3}. 
Combining \eqref{Eq:Proof:d3} and \eqref{Eq:Proof:d2} gives the desired result. \hfill $\blacksquare$

\section{Proof of Lemma~\ref{Lem:CCOPE:LB:Rich}}
\label{App:CCOPE:LB:Rich}
To apply Lemma~\ref{Lem:Sum:RV}, define $Z'_n = Z_n/\ell^{\frac{1}{\nu}}$ and $z = \ell^{\frac{\alpha}{2}-\frac{1}{\nu}} x$ and rewrite \eqref{Eq:Zn:Rich} as 
\begin{align}
\Pr(Z'_n > z) &\sim \frac{\omega^N\Gamma\l(\frac{\alpha \nu}{2} + 1\r)\Pr(N=\nu)}{(\pi\lambda)^{\frac{\alpha \nu}{2}}} z^{-\nu} \label{Eq:Zn:Rich:a}	
\end{align}
as $z \rightarrow\infty$. It can be observed from \eqref{Eq:Zn:Rich:a} that $Z'_n\in \mathsf{RV}(\nu)$. Moreover, given $\alpha > 2$, the sum $\sum_{n=1}^\infty n^{-\frac{\alpha}{2}}$ can be checked to be finite.  Therefore,  using \eqref{Eq:Zn:Rich:a} and applying Lemma~\ref{Lem:Sum:RV},  
\begin{equation}
\begin{aligned}
\!\! \Pr\l(\sum_{n=1}^\infty n^{-\frac{\alpha}{2}} Z'_n > z \r) \sim &\frac{\omega^N\Gamma\l(\frac{\alpha \nu}{2} + 1\r)\Pr(N=\nu)}{(\pi\lambda)^{\frac{\alpha \nu}{2}}} \times \\
 &\qquad \qquad  z^{-\nu}\sum_{n=1}^\infty n^{-\frac{\alpha}{2}}. 
 \end{aligned}\label{Eq:Z:Tail:Rich}
\end{equation}
Combining  the definitions of $Z'_n$ and $z$ and  \eqref{Eq:Z:Tail:Rich} yields 
\begin{equation}
\begin{aligned}
 \!\!\!\!\Pr\l(\sum_{n=1}^\infty n^{-\frac{\alpha}{2}} Z_n >\ell^{\frac{\alpha}{2}} x\r) \sim &\frac{\omega^N\Gamma\l(\frac{\alpha \nu}{2} + 1\r)\Pr(N=\nu)}{(\pi\lambda)^{\frac{\alpha\nu}{2}}} \\
&\ell^{-(\frac{1}{2} \alpha \nu - 1)} x^{-\nu}\sum_{n=1}^\infty n^{-\frac{\alpha}{2}}. 
 \end{aligned}\label{Eq:Z:Tail:Rich:a}
\end{equation}
The desired asymptotic lower bound on the OPE follows from   \eqref{Eq:CCOPE:LB:Sparse}   and \eqref{Eq:Z:Tail:Rich:a}. \hfill $\blacksquare$

\bibliographystyle{ieeetr}

\begin{IEEEbiographynophoto}{Kaibin Huang}  (S'05, M'08) received the B.Eng. (first-class hons.) and the M.Eng. from the National University of Singapore in 1998 and 2000, respectively, and the Ph.D. degree from The University of Texas at Austin (UT Austin) in 2008, all in electrical engineering.

Since Jul. 2012, he has been an assistant professor in the Dept. of Applied Mathematics at The Hong Kong Polytechnic University, Hong Kong. He had held the same position in the School of Electrical and Electronic Engineering at Yonsei University, S. Korea from Mar. 2009 to Jun. 2012 and presently is affiliated with the school as an adjunct professor. From Jun. 2008 to Feb. 2009, he was a Postdoctoral Research Fellow in the Department of Electrical and Computer Engineering at the Hong Kong University of Science and Technology. From Nov. 1999 to Jul. 2004, he was an Associate Scientist at the Institute for Infocomm Research in Singapore. He frequently serves on the technical program committees of major IEEE conferences in wireless communications. He will chair the Comm. Theory Symp. of IEEE ICC 2014 and has been the technical co-chair for IEEE CTW 2013, the track chair for IEEE Asilomar 2011, and the track co-chair for IEE VTC Spring 2013 and IEEE WCNC 2011. He is an editor for the IEEE Wireless Communications Letters and also the Journal of Communication and Networks. He is an elected member of the SPCOM Technical Committee of the IEEE Signal Processing Society. Dr. Huang received the Outstanding Teaching Award from Yonsei, Motorola Partnerships in Research Grant, the University Continuing Fellowship at UT Austin, and a Best Paper award at IEEE GLOBECOM 2006. His research interests focus on the analysis and design of wireless networks using stochastic geometry and multi-antenna limited feedback techniques.
\end{IEEEbiographynophoto}

\begin{IEEEbiographynophoto}
{Jeffrey Andrews}(SÕ98, MÕ02, SMÕ06, FÕ13) received the B.S. in Engineering with High Distinction from Harvey Mudd College in 1995, and the M.S. and Ph.D. in Electrical Engineering from Stanford University in 1999 and 2002, respectively.  He is a Professor in the Department of Electrical and Computer Engineering at the University of Texas at Austin, where he was the Director of the Wireless Networking and Communications Group (WNCG) from 2008-12. He developed Code Division Multiple Access systems at Qualcomm from 1995-97, and has consulted for entities including the WiMAX Forum, Intel, Microsoft, Apple, Clearwire, Palm, Sprint, ADC, and NASA. 

Dr. Andrews is co-author of two books, Fundamentals of WiMAX (Prentice-Hall, 2007) and Fundamentals of LTE (Prentice-Hall, 2010), and holds the Earl and Margaret Brasfield Endowed Fellowship in Engineering at UT Austin, where he received the ECE departmentÕs first annual High Gain award for excellence in research. He is a Senior Member of the IEEE, a Distinguished Lecturer for the IEEE Vehicular Technology Society, served as an associate editor for the IEEE Transactions on Wireless Communications from 2004-08, was the Chair of the 2010 IEEE Communication Theory Workshop, and is the Technical Program co-Chair of ICC 2012 (Comm. Theory Symposium) and Globecom 2014.  He is an elected member of the Board of Governors of the IEEE Information Theory Society and an IEEE Fellow.
 
Dr. Andrews received the National Science Foundation CAREER award in 2007 and has been co-author of five best paper award recipients, two at Globecom (2006 and 2009), Asilomar (2008), the 2010 IEEE Communications Society Best Tutorial Paper Award, and the 2011 Communications Society Heinrich Hertz Prize.  His research interests are in communication theory, information theory, and stochastic geometry applied to wireless cellular and ad hoc networks.
\end{IEEEbiographynophoto}
\vfill 

\end{document}